\documentclass{aa}

\usepackage{graphics}
\usepackage{epsfig}

\def\bib{\bibitem{}}

\newcommand{\rhoa}{\overline{\rho}}
\newcommand{\gam}{\gamma}

\newcommand{\beq}{\begin{equation}}
\newcommand{\eeq}{\end{equation}}
\newcommand{\lag}{\langle}
\newcommand{\rag}{\rangle}
\newcommand{\Om}{\Omega_{\rm m}}
\newcommand{\Ol}{\Omega_{\Lambda}}
\newcommand{\Ob}{\Omega_{\rm b}}
\newcommand{\Omo}{\Omega_{\rm m,0}}
\newcommand{\De}{{\cal D}}
\newcommand{\om}{\omega}
\newcommand{\hgam}{\hat{\vec \gam}}
\newcommand{\na}{\overline{n}}
\newcommand{\xiIGM}{\xi_{\rm IGM}}
\newcommand{\cP}{{\cal P}}
\newcommand{\Vion}{V_{\rm ion}}
\newcommand{\xcool}{x_{\rm cool}}
\newcommand{\CIGM}{C_{\rm IGM}}

\newcommand{\TIGM}{T_{\rm IGM}}
\newcommand{\Rd}{R_{\rm d}}
\renewcommand{\d}{{\rm d}}
\newcommand{\xie}{\xi_e}

\newcommand{\xieu}{\xi_{e,{\rm u}}}
\newcommand{\xieh}{\xi_{e,{\rm h}}}
\newcommand{\xiegal}{\xi_{e,{\rm gal}}}
\newcommand{\xigal}{\xi_{\rm gal}}
\newcommand{\Fgalvol}{F_{\rm gal,vol}}
\newcommand{\Fgal}{F_{\rm gal}}
\newcommand{\Cgal}{C_{\rm gal}}
\newcommand{\Rcool}{R_{\rm cool}}
\newcommand{\zri}{z_{\rm ri}}
\newcommand{\Cs}{C_{\rm s}}
\newcommand{\CQ}{C_{\rm Q}}
\newcommand{\Cu}{C_{\rm u}}
\newcommand{\Ch}{C_{\rm h}}
\newcommand{\Sl}{S_l}
\newcommand{\Slgal}{S_{l,{\rm gal}}}
\newcommand{\Slu}{S_{l,{\rm u}}}
\newcommand{\Slh}{S_{l,{\rm h}}}
\newcommand{\Chs}{C_{\rm h,s}}
\newcommand{\ChQ}{C_{\rm h,Q}}
\newcommand{\Sls}{S_{l,{\rm s}}}
\newcommand{\SlQ}{S_{l,{\rm Q}}}
\newcommand{\Slhs}{S_{l,{\rm h,s}}}
\newcommand{\SlhQ}{S_{l,{\rm h,Q}}}
\newcommand{\Cl}{C_l}
\newcommand{\Ctheta}{C_{\theta}}
\newcommand{\Cthetagal}{C_{\theta,{\rm gal}}}
\begin{document}
%
% US additions
% 
%\topmargin=2.5 cm
%\evensidemargin=2.5 cm
%\oddsidemargin=2.5 cm
%\thispagestyle{empty}
%
%
\renewcommand{\textfraction}{.01}
\renewcommand{\topfraction}{0.99}
\renewcommand{\bottomfraction}{0.99}
\setlength{\textfloatsep}{2.5ex}
\thesaurus{Sect.02 (12.03.1; 12.03.4; 11.09.3; 12.12.1)}
\title{Secondary CMB anisotropies from the kinetic SZ effect}   
\author{P. Valageas\inst{1}, A. Balbi\inst{2} and J.
Silk\inst{3,}\inst{4}}
\institute{Service de Physique Th\'eorique, CEA Saclay, 91191
Gif-sur-Yvette, 
France
\and
Dipartimento di Fisica, Universit\`a Tor Vergata, Roma I-00133, Italy
\and
Astrophysics, Department of Physics, Keble Road, Oxford OX1 3RH, U.K.
\and
Institut d'Astrophysique de Paris, CNRS, 98bis Boulevard Arago,
F-75014 Paris, France}
\date{Received / Accepted }
\maketitle
\markboth{P. Valageas, A. Balbi \& J. Silk: Secondary CMB anisotropies
from the 
kinetic SZ effect}{P. Valageas, A. Balbi \& J. Silk: Secondary CMB
anisotropies 
from the kinetic SZ effect}

\begin{abstract}

The reionization of the universe by stars and quasars is expected to be
a highly 
inhomogeneous process. Moreover, the fluctuations of the matter density
field 
also lead to inhomogeneities of the free electron distribution. These
patterns 
gave rise to secondary CMB anisotropies through Thomson scattering of
photons by 
free electrons. In this article we present an analytic model, based on
our 
previous work which tackled the reionization history of the universe,
which 
allows us to describe the generation of these secondary CMB
anisotropies. We 
take into account the ``patchy pattern'' of reionization (HII bubbles),
the 
cross-correlations of these ionized regions, the small-scale
fluctuations of the 
matter density field and the contribution from collapsed objects.

For an open universe, we find that the angular correlation function
$C(\theta)$ 
displays a very slow decline from $C(0) \sim 6 \times 10^{-13}$ up to
the scale 
$\theta \sim 10^{-3}$ rad where it shows a sharp drop. On the other
hand, the 
power-spectrum $l(l+1)\Cl/(2\pi)$ (and the ``local average'' $\Sl$)
exhibits a 
plateau of height $\sim 10^{-13}$ in the range $10^3 < l < 10^6$. We
find that 
for large wavenumbers $l > 10^4$ the signal is dominated by the
contribution 
from collapsed halos while for $l < 10^4$ it is governed by the
large-scale 
correlations of HII bubbles. This implies that one cannot discriminate 
reionization by stars from a quasar-driven scenario since the size of
ionized 
regions never dominates the behaviour of the anisotropies. Moreover, the 
secondary CMB anisotropies arise from a broad range of redshifts ($7.5 <
z <10$ 
for the IGM and $0<z<7$ for galactic halos). Thus, we find that the
generation 
of these anisotropies involves several intricate processes and they are
close to 
the resolution limit of current numerical simulations.
The signal expected in our model might bias the cosmological
parameter estimation from CMB experiments such as Planck or MAP, and
could be detected by future mm-wavelength interferometers (e.g., ALMA).

\end{abstract}


\section{Introduction}

Observations of the spectra of distant quasars show that the universe is
highly 
ionized by $z=5$, while recombination took place at $z \sim 1100$. In
current 
cosmological scenarios, the reionization (and reheating) of the universe
occurs 
at $\zri \sim 10$ (typically $6 \la \zri \la 15$) when structure
formation is 
sufficiently advanced to build a large number of radiation sources
(galaxies or 
quasars) which photoionize the IGM (e.g., Valageas \& Silk 1999a).
However, the 
whole reionization history is a gradual and inhomogeneous process: each
emitting 
object builds an HII region in its surroundings and reionization occurs
when 
these bubbles overlap. This last stage is very rapid (e.g., Gnedin 2000)
but at 
earlier redshifts there is a very inhomogeneous phase which evolves
rather 
slowly, as the size of the ionized regions grows and the number of
radiation 
sources increase. Then, this process can leave an imprint on the CMB
through 
Thomson scattering of photons from free electrons. First, the mixing of
photons 
coming from different lines of sight leads to a damping of small-scale
primary 
fluctuations. Second, the Doppler effect (photons get some of the
peculiar 
momentum of free electrons) generates secondary anisotropies since the 
distribution of free electrons is highly inhomogeneous. Thus,
observations of 
CMB anisotropies could provide some information on the properties of the 
reionization process and on the features of the IGM at high redshifts.

As pointed out by Sunyaev (1978) and Kaiser (1984) the oscillations of
the 
velocity field (as opposite sides of overdensities have almost opposite 
velocities) lead to a strong suppression of these secondary
anisotropies. 
However, the modulation produced by the spatial variation of the number
density 
of free electrons removes this cancelation on small scales and can
generate 
significant CMB anisotropies. These fluctuations of the density of free 
electrons can be produced by several processes. First, as explained
above, 
spatial variations of the ionized fraction of hydrogen due to patchy 
reionization provide a source of inhomogeneities (even if the IGM is
uniform). 
This is relevant before reionization. Second, the fluctuations of the
matter 
density field itself lead to inhomogeneities of the density of free
electrons. 
This occurs both before and after reionization. When the density
fluctuations 
are in the linear regime this corresponds to the Ostriker-Vishniac
effect 
(Ostriker \& Vishniac 1986) while the non-linear regime is usually
called the 
kinetic Sunyaev-Zel'dovich effect (e.g., Sunyaev \& Zel'dovich 1980). 

In this article, we study both processes (patchy reionization and matter
density 
fluctuations) in a unified fashion. To this order, we use an analytic
model 
described in a previous paper (Valageas \& Silk 1999a) which we built to 
investigate the reionization and reheating history of the universe. It
includes 
a model for galaxy formation (described in details in Valageas \&
Schaeffer 
1999) and for the quasar multiplicity function, which have been compared
with 
observations at low redshifts ($z < 4.5$). Moreover, it also provides a 
description of the correlations of the matter density field which is
consistent 
with these mass functions. The underlying model of the non-linear
density field 
is based on the stable-clustering ansatz as detailed in Balian \&
Schaeffer 
(1989) (see also Valageas \& Schaeffer 1997). This allows us to take
into 
account density fluctuations within the IGM, the reionization process
through 
the creation of HII regions and the correlations of these ionized
bubbles. In 
addition to the IGM, we also consider the contribution from galactic
halos. Here 
we restrict ourselves to the temperature anisotropies and we do not
consider 
polarization. Thus, the main goals of this article are to:

- present an analytic model which can describe in a more detailed
fashion than 
previous works the generation of these secondary CMB anisotropies.
Moreover, our 
approach is self-consistent and it agrees with observations at low
redshifts.

- obtain the redshift distribution of the contributions to these
anisotropies.

- take advantage of the fact that we use an analytic model to explicitly 
separate the contributions from different physical processes
(small-scale 
density fluctuations, patchy reionization, correlations of ionized
bubbles). 
This allows us to see which information can be recovered from
observations of 
the CMB.

- investigate whether one can discriminate reionization by stars versus
quasars 
from the observed CMB anisotropies.

- check whether the main features of this process strongly depend on the 
cosmological scenario.

We present in Sect.\ref{Secondary anisotropies} the formalism we use to
compute 
these secondary CMB anisotropies, as well as the approximations we
introduce. 
Then, in Sect.\ref{Numerical results} we describe our numerical results
for the 
case of an open universe. This corresponds to the model we used in a
previous 
work (Valageas \& Silk 1999a) to investigate the reionization history of
the 
universe. In the last section we also consider briefly a critical
density 
universe for comparison.

\section{Secondary anisotropies}
\label{Secondary anisotropies}

\subsection{Contribution from inhomogeneities of the free electron
distribution}
\label{Contribution from inhomogeneities of the free electron
distribution}

As described for instance in Gruzinov \& Hu (1998) and Knox et al.
(1998), 
Thomson scattering of CMB photons off moving free electrons in the IGM
generates 
secondary anisotropies. Thus, for small optical depths the temperature 
perturbation $\Delta_T(\hgam) = \delta T /T$ on the direction $\hgam$ on
the sky 
is:
\begin{eqnarray} 
{\displaystyle  \Delta_T(\hgam) = - \int \d {\bf l} . \frac{\bf v}{c} \; \sigma_T n_{e,\rm f} e^{-\tau} = - \frac{\sigma_T c}{H_0} \int \frac{\d\chi}{1+z} \; \frac{\hgam .\bf v}{c} n_{e,\rm f} } \nonumber \\
\label{DeltaT}
\end{eqnarray}
where $\sigma_T$ is the Thomson cross-section, $n_{e,\rm f}$ the number
density 
of free electrons, ${\bf v}$ the peculiar velocity and $l$ the
coordinate along 
the line of sight, all in physical units. We also defined $\chi$ as the 
dimensionless radial comoving coordinate:
\beq
\d\chi = \frac{\d z}{\sqrt{\Ol+(1-\Om-\Ol)(1+z)^2+\Om(1+z)^3}} .
\label{chi}
\eeq
Since in our scenario reionization occurs at $z=6.8$, the optical depth $\tau$ from $\chi=0$ up to $\chi$ in (\ref{DeltaT}) is very small ($\tau < 0.024$, see Fig. 15 in Valageas \& Silk 1999a) and it plays no role so that we used the approximation $e^{-\tau} \simeq 1$.
Then, the two-point correlation function $C(\theta)$ of these
temperature 
distortions is simply:
\beq
\begin{array}{l} {\displaystyle C(\theta) = \left( \frac{\sigma_T \;
c}{H_0} 
\right)^2 \int \frac{\d\chi_1}{1+z_1} \int \frac{\d\chi_2}{1+z_2} } \\
\\ 
{\displaystyle \hspace{2.5cm} \times \; \lag \frac{\hgam_1 .{\bf
v}_1}{c} \; 
\frac{\hgam_2 .{\bf v}_2}{c} \;\; n_{e,\rm f,1} \; n_{e,\rm f,2} \rag } 
\end{array}
\label{Corr1}
\eeq
where the directions $\hgam_1$ and $\hgam_2$ make the angle $\theta$.
One can 
distinguish two effects which contribute to the correlation function in 
(\ref{Corr1}). Firstly, a uniform reionization in a homogeneous IGM
(i.e. 
$n_{e,\rm f,1}$ shows no fluctuations) provides a non-zero contribution
through 
the fluctuations $\lag (\hgam_1 . {\bf v}_1) (\hgam_2 . {\bf v}_2) \rag$
of the 
velocity field. However, because of the oscillations of the velocity
correlation 
(related to the infall from opposite sides into potential wells) the
integration 
along the line of sight leads to a strong suppression of these Doppler
effects 
(Sunyaev 1978; Kaiser 1984). Secondly, the inhomogeneities of the free
electron 
number density add a second contribution which avoids this cancelation
of the 
velocity term. Most of the power which builds the fluctuations of the
velocity 
comes from scales of the order of $R_{-1}$ where the local slope of the 
power-spectrum is $n=-1$. This gives $R_{-1} \sim 10$ comoving Mpc. On
the other 
hand, the characteristic scale of the inhomogeneities of the electron 
distribution is of order $R_{n_e} \sim 100$ comoving kpc (it is set by
the size 
of the ionized bubbles and the scale of non-linear structures). Thus, we
make 
the approximation that the velocity fluctuations are not correlated to
the 
density field and we write:
\begin{eqnarray} 
{\displaystyle C(\theta) = \tau_0^2 \int \d\chi_1 \d\chi_2 \; (1+z_1)^2 
(1+z_2)^2 \; \lag \frac{\hgam_1 .{\bf v}_1}{c} \; \frac{\hgam_2 .{\bf
v}_2}{c} 
\rag } \nonumber \\  \nonumber \\  {\displaystyle \hspace{0.cm} \times
\left[ 
\lag (1+\delta_1) x_{e,1} (1+\delta_2) x_{e,2} \rag - \lag (1+\delta_1)
x_{e,1} 
\rag \lag (1+\delta_2) x_{e,2} \rag \right] } \nonumber \\
\label{Corr2}
\end{eqnarray}
where $x_e = n_{e,\rm f}/n_e$ is the ionization fraction, $\delta = (n_b
- 
\na_b)/\na_b$ is the density contrast and we defined:
\beq
\tau_0 = \frac{n_{e,0} \; \sigma_T \; c}{H_0} .
\label{tau0}
\eeq
Here $n_{e,0}$ is the present number density of free electrons:
\beq
n_{e,0} = \frac{\Omega_b}{\Om} \; \frac{\rhoa_0}{m_p} \; \left( 1 -
\frac{Y}{2} 
\right)
\eeq
where $Y=0.26$ is the helium mass fraction. The difference which enters
the term 
in the second line of expression (\ref{Corr2}) corresponds to the fact
that here 
we only consider the contribution to secondary CMB anisotropies due to
the 
modulation of the velocity field by the fluctuations of the free
electron number 
density. Thus, for a uniform free electron density field the quantity 
$C(\theta)$ written in (\ref{Corr2}) vanishes. As we discussed above,
the 
factorization of the velocity and density averages we used in
(\ref{Corr2}) is 
valid at small scales $r \ll R_{-1}$ where we actually have: $\lag
(\hgam_1 
.{\bf v}_1) (\hgam_2 .{\bf v}_2) \rag \simeq \lag (\hgam_1 .{\bf v}_1)
(\hgam_2 
.{\bf v}_1) \rag \simeq \lag (\hgam_1 .{\bf v}_1)^2 \rag$. To get the
last 
approximation we used the fact that the lines of sight $\hgam_1$ and
$\hgam_2$ 
must be nearly parallel since all scales of interest are much smaller
than the 
Hubble scale. In other words, at small distances we can consider the
velocity 
field to be constant. However, in practice we shall also consider larger
scales 
where the cross-correlation of ionized bubbles generates some power.
These 
scales are close to $R_{-1}$ hence we keep the expression (\ref{Corr2})
rather 
than making the approximation $\lag (\hgam_1 .{\bf v}_1) (\hgam_2 .{\bf
v}_2) 
\rag \simeq \lag v^2 \rag/3$. However, on these scales we can expect to 
underestimate somewhat the angular correlation $C(\theta)$ since the
velocity 
field should be correlated with the density field. We note $\xi_v({\bf 
R}_{\theta} + {\bf l})$ the velocity term and $\xi_e({\bf R}_{\theta} +
{\bf 
l})$ the electron density correlation which enter the expression
(\ref{Corr2}) 
and we write (\ref{Corr2}) as:
\beq
C(\theta) = \tau_0^2 \int \d\chi \; \frac{H_0}{c} (1+z)^5 
\int_{-\infty}^{\infty} \d l \; \xi_v({\bf R}_{\theta} + {\bf l}) \;
\xi_e( {\bf 
R}_{\theta} + {\bf l} ) .
\label{Corr3}
\eeq
Here $l$ is the physical length along the line of sight while ${\bf
R}_{\theta}$ 
is the physical distance at redshift $z$ between both lines of sight
(${\bf 
R}_{\theta}$ is orthogonal to ${\bf l}$):
\beq
R_{\theta}(z) = \frac{c}{H_0} \; \frac{\De(z)}{1+z} \; \theta
\label{Rtheta}
\eeq
where the dimensionless comoving angular distance $\De$ is defined by:
\beq
\De(z) = \frac{1}{\sqrt{1-\Om-\Ol}} \sinh \left( \sqrt{1-\Om-\Ol} \;\;
\chi 
\right) .
\label{De}
\eeq
In this article we only consider the secondary anisotropies due to the 
inhomogeneities of the free electron number distribution, described by 
(\ref{Corr3}). As discussed above, the contribution which arises from
the 
fluctuations of the velocity field only is smaller because of the
cancelations 
along the line of sight.

\subsection{Velocity fluctuations}
\label{Velocity fluctuations}

First, the velocity term in (\ref{Corr3}) is obtained as follows. Since
most of 
the scales which contribute to the velocity fluctuations are in the
linear 
regime until $z=0$, we can use the linear relation between the velocity
and 
density fluctuations (Peebles 1980):
\beq
{\bf v}({\bf x}) = a H(z) f[\Om(z)] \int \d^3k \; e^{i {\bf k . x}} \;
\frac{i 
{\bf k}}{k^2} \; \delta({\bf k})
\eeq
where $f(\Om) \simeq \Om^{0.6}$ if $\Ol=0$, $a$ is the scale factor
($a=1$ at 
$z=0$), ${\bf k}$ and ${\bf x}$ are comoving quantities. Since we
consider lines 
of sight which are almost parallel (small angles $\theta$) we write the
velocity 
correlation as (e.g., Groth et al. 1989):
\begin{eqnarray}
\xi_v({\bf x}) & = & \frac{4\pi}{c^2} \left( a H(z) f[\Om(z)] \right)^2 
\int_0^{\infty} \d k \; P(k;z) \nonumber \\ \nonumber \\ & &
\hspace{2cm} \times 
\left[ j_0(k x) - 2 \; \frac{j_1(k x)}{k x} \right]
\end{eqnarray}
where $j_0$ and $j_1$ are the spherical Bessel functions. Here we
introduced the 
linear power-spectrum defined by $\lag \delta({\bf k}_1) \delta({\bf
k}_2) \rag 
= P(k_1) \delta_D({\bf k}_1+{\bf k}_2)$. Next we define at $z=0$:
\beq
v_0^2 = 4\pi \; \left( H_0 f(\Omo) \right)^2 \; \int_0^{\infty} \d k \;
P(k;0)
\eeq
and
\beq
\Pi_v(x) = 3 \; \frac{ \int \d k \; P(k;0) \; \left[ j_0(k x) - 2 \;
\frac{j_1(k 
x)}{k x} \right] } { \int \d k \; P(k;0) } .
\eeq
Thus, $\Pi_v(x)$ only depends on the shape of the power-spectrum and we
have 
$\Pi_v(0)=1$ while at large scales $x \gg R_{-1}$ it shows an
oscillatory 
behaviour. Then, we can write $\xi_v({\bf x})$ as:
\beq
\xi_v({\bf x}) = \frac{v_0^2}{3 c^2} \; \Pi_v(x) \; \left[ (1+z)^{-2} 
\frac{H(z)}{H_0} \frac{f[\Om(z)]}{f(\Omo)} \right]^2
\label{v1}
\eeq
where we used the redshift evolution $P(k) \propto (1+z)^{-2}$ of the
linear 
power-spectrum.

\subsection{Free electron density fluctuations}
\label{Free electron density fluctuations}

Second, we have to model the fluctuations of the free electron density 
distribution. As seen in (\ref{Corr2}) two effects contribute to these 
fluctuations: i) inhomogeneities of the baryonic matter distribution and
ii) of 
the ionization fraction. In most previous studies of these secondary CMB 
anisotropies only the second contribution was taken into account through
a model 
of ionized bubbles within a uniform IGM (e.g., Gruzinov \& Hu 1998; Knox
et al. 
1998). However, as shown in Valageas \& Silk (1999a) the clumping of the
gas is 
not negligible even at $z \sim 10$. As can be seen from (\ref{Corr2}),
and as we 
shall check below, this increases somewhat the amplitude of these CMB 
anisotropies. In this article we use a simple model to estimate the
correlation 
term $\xi_e(r)$. First, although the baryonic density fluctuations may
be 
correlated with the ionization fraction we make the approximation:
\beq
\xi_e \simeq \lag (1+\delta_1) (1+\delta_2) \rag \lag x_{e,1} x_{e,2}
\rag - 
\lag 1+\delta \rag^2 \lag x_e \rag^2
\label{xie0}
\eeq
which can be written:
\beq
\! \xi_e(r) = \left( [1+\xiIGM(r)] \lag x_{e,1} x_{e,2} \rag(r) - \lag
x_e 
\rag^2 \right) \lag 1+\delta \rag_{\rm IGM}^2 .
\label{xie1}
\eeq
Here we note $\xiIGM(r)$ the two-point correlation function of the gas
density 
within the IGM and $\lag \delta \rag_{\rm IGM}$ the mean density
contrast of the 
IGM. Note that at low redshifts after reionization $x_e=1$ almost
everywhere 
within the IGM and the relation $\xi_e(r) = \xiIGM(r) \lag 1+\delta
\rag_{\rm 
IGM}^2$ becomes exact. Until $z \sim 1$ most of the matter is contained
in the 
IGM since only a very small amount of gas was able to cool and form
galaxies, 
hence $\lag 1+\delta \rag_{\rm IGM}$ is close to unity, while at $z=0$
we have 
$\lag 1+\delta \rag_{\rm IGM} \sim 0.4$ (Valageas \& Silk 1999a;
Valageas et al. 
2000).

\subsubsection{Ionization fraction}
\label{Ionization fraction}

In order to obtain the term $\lag x_{e,1} x_{e,2} \rag(r)$ we use a
model of spherical ionized bubbles around galaxies and quasars as in Valageas \& Silk
(1999a). 
Thus, we consider that $x_e=1$ within ionized patches and $x_e=0$
everywhere 
else. At low $z$ after overlap of the ionized regions $x_e=1$ throughout
the 
IGM. Indeed, reionization occurs thanks to the growth of the ionized
bubbles 
which finally occupy all the volume (and not through a slow increase of
a 
uniform ionization fraction). Then, $\lag x_{e,1} x_{e,2} \rag(r)$ is
simply the 
probability that two points at distance $r$ are located within ionized
regions. 
First, let us consider uncorrelated ionizing sources. Then, the
probability that 
the first point ${\bf r}_1$ is within an ionized bubble is the volume
fraction 
occupied by these regions:
\beq
\cP_1 = \int \frac{\d x_1}{x_1} \frac{\rhoa}{M_1} x_1^2 H(x_1)
\Vion(x_1)
\label{Proba1}
\eeq
where $\Vion(x_1)$ is the volume of the ionized bubble associated to the
source 
$x_1$ (galaxy or quasar) and $(\rhoa/M_1) x_1 H(x_1) \d x_1$ is the number density of radiation sources labelled by the parameter $x_1$ in the range $[x_1 , x_1+\d x_1]$ (see Valageas \& Silk 1999a; Valageas \& Schaeffer 1997). Next, a simple geometrical calculation shows
that the 
second point ${\bf r}_2={\bf r}_1+{\bf r}$ can be located within the
same 
ionized spherical bubble of radius $R_1$ with a probability $\cP_{2,\rm
s}$:
\beq
\cP_{2,\rm s} = F\left( \frac{r}{R_1} \right) \hspace{0.2cm} \mbox{with}
\left\{ 
\begin{array}{l} F(x)= 1- \frac{3x}{4}+\frac{x^3}{16} \;\;\; \mbox{if}
\; x<2 \\ 
\\ F(x)=0 \;\;\; \mbox{if} \; x \geq 2 \end{array} \right.
\eeq
On the other hand, the point ${\bf r}_2$ can be embedded within another
bubble 
with a probability $\cP_{2,\rm o}$ with:
\beq
\cP_{2,\rm o} = \left[ 1 - F\left( \frac{r}{R_1} \right) \right] \int
\frac{\d 
x_2}{x_2} \frac{\rhoa}{M_2} x_2^2 H(x_2) \Vion(x_2)
\eeq
where we assumed that the bubbles associated to different sources do not 
overlap. This should be a good approximation before reionization when
there are 
very few ionized regions. Hence we obtain for uncorrelated sources $\lag
x_{e,1} 
x_{e,2} \rag(r) = \cP_1 (\cP_{2,\rm s}+\cP_{2,\rm o})$. However, the
sources 
$x_1$ and $x_2$ should be correlated, especially at high $z$ where they 
correspond to very rare large density fluctuations. Thus, we write:
\beq
\begin{array}{l} {\displaystyle \cP_2(r;x_1) = \mbox{Min} \Biggl \lbrace
1 \; ,  
\; F\left( \frac{r}{R_1} \right) \; + \; \left[ 1-F\left( \frac{r}{R_1}
\right) 
\right] } \\ \\ {\displaystyle \hspace{1cm} \times \int \frac{\d
x_2}{x_2} 
\frac{\rhoa}{M_2} x_2^2 H(x_2) \Vion(x_2) \left[ 1+\xi_{x_1,x_2}(r)
\right]
 \Biggl \rbrace } \end{array}
\label{P2}
\eeq
where $\xi_{x_1,x_2}(r)$ is the correlation function of the objects
$x_1$ and 
$x_2$. Following the analytic results obtained in Bernardeau \&
Schaeffer (1992, 
1999) we have:
\beq
\xi_{x_1,x_2}(r) = b(x_1) b(x_2) \xi(r)
\label{bias}
\eeq
with
\beq
x \ll 1 : \; b(x) \propto x^{(1-\om)/2} \hspace{0.5cm} \mbox{and}
\hspace{0.5cm} 
x \gg 1 : \; b(x) \propto x
\eeq
where $b(x)$ is the bias associated to an object defined by the
parameter $x$. 
This behaviour has been shown to agree with the results of numerical
simulations 
in Munshi et al. (1999). In Valageas et al. (2000) the predictions of
this model 
for the correlation functions of galaxies, quasars, Lyman-$\alpha$
clouds and 
clusters have been compared to observations. Thus, this provides a
unified model 
for all these objects (which are characterized by different scales and 
densities) which agrees reasonably well with observations and
simulations. In 
(\ref{P2}) the use of the minimum ensures that $\cP_2 \leq 1$. Indeed,
at small 
redshift after reionization the second term in the minimum of (\ref{P2})
becomes 
larger than unity since all the volume is ionized (the ionized bubbles
overlap 
and $\Vion$ goes to infinity). Finally, we obtain $\lag x_{e,1} x_{e,2}
\rag(r)$ 
as:
\beq
\begin{array}{l} {\displaystyle \lag x_{e,1} x_{e,2} \rag(r) = } \\ \\ 
{\displaystyle \hspace{0.9cm} \mbox{Min} \left\{ 1 \; , \; \int \frac{\d 
x_1}{x_1} \frac{\rhoa}{M_1} x_1^2 H(x_1) \Vion(x_1) \cP_2(r;x_1)
\right\} } 
\end{array}
\label{Xixe}
\eeq
while the mean ionization fraction is simply:
\beq
\lag x_e \rag = \mbox{Min} \left\{ 1 \; , \; \int \frac{\d x_1}{x_1} 
\frac{\rhoa}{M_1} x_1^2 H(x_1) \Vion(x_1) \right\} .
\label{xe}
\eeq
The minima in (\ref{Xixe}) and (\ref{xe}) ensure again that $\lag
x_{e,1} 
x_{e,2} \rag \leq 1$ and $\lag x_e \rag \leq 1$ after reionization.

\subsubsection{Density fluctuations}
\label{Density fluctuations}

Finally, we need to evaluate the two-point correlation function
$\xiIGM(r)$ of 
the gas density within the IGM which appears in (\ref{xie1}). We
consider that 
the universe is made of collapsed objects which have been able to cool
and to 
form galaxies, embedded within a lower density medium which we call the
IGM. 
Hence, the latter corresponds to voids as well as to density
fluctuations (which 
may appear as filaments or shallow spherical halos) associated with the 
Lyman-$\alpha$ forest (Valageas et al. 1999). Then, as in Valageas \&
Silk 
(1999a) the mean density of the matter which forms the IGM is given by:
\beq
\lag 1+\delta \rag_{\rm IGM} = (1+\Delta)_u + \int_0^{\xcool} \frac{\d
x}{x} \; 
x^2 H(x)
\label{DeltaIGM}
\eeq
while the mean square density is:
\beq
\lag (1+\delta)^2 \rag_{\rm IGM} = (1+\Delta)_u^2 + \int_0^{\xcool}
\frac{\d 
x}{x} \; x^2 H(x) (1+\Delta) .
\label{Delta2IGM}
\eeq
The term $(1+\Delta)_u$ corresponds to voids while the second term in
the r.h.s. 
in (\ref{DeltaIGM}) and (\ref{Delta2IGM}) is the contribution from 
Lyman-$\alpha$ forest clouds. The upper bound $\xcool$ ensures that we
do not 
count cooled objects (which are identified to galactic halos) nor
clusters of 
galaxies, since they are not part of the IGM. Then, the clumping factor
of the 
IGM is simply:
\beq
\CIGM \equiv 1 + \xiIGM(r=0) = \frac{ \lag (1+\delta)^2 \rag_{\rm IGM} }
{ \lag 
1+\delta \rag_{\rm IGM}^2 }
\label{CIGM}
\eeq
(which we noted $C_n$ in Valageas \& Silk 1999a). In addition, as
described in 
Valageas et al. (1999) and Valageas \& Silk (1999a), baryonic density 
fluctuations within the IGM are erased on small scales below $\Rd(z)$
due to the 
non-zero temperature $\TIGM$ with:
\beq
\Rd(z) \sim t_H \sqrt{ \frac{k \TIGM}{\mu m_p} } \sim t_H \; c_s .
\label{Rd}
\eeq
This corresponds to the scale reached by acoustic waves over the Hubble
time 
$t_H$ (with the sound velocity $c_s$). Note that the damping scale $\Rd$
is 
usually smaller than the standard Jeans scale $R_{\rm Jeans}$ which
corresponds 
to the limit of large times. Hence we write:
\beq
r \leq \Rd(z) : \hspace{0.2cm} \xiIGM(r) = \xiIGM(\Rd) = \CIGM - 1
\label{cutRd}
\eeq
in order to take into account the damping of the baryonic power-spectrum
on 
small scales within the IGM, in a fashion which is consistent with
(\ref{CIGM}). 
Finally, for larger scales we use the prescription:
\beq
r > \Rd(z) : \hspace{0.2cm} \xiIGM(r) = (\CIGM-1) \;
\frac{\xi(r)}{\xi(\Rd)} .
\label{xiGM}
\eeq
Note that this large$-r$ behaviour (i.e. for $r > \Rd$) is consistent
with the 
factorization (\ref{bias}).

\subsection{Contribution from galactic halos}
\label{Contribution from galactic halos}

In addition to the free electron IGM number density correlation function 
$\xi_e(r)$ which we defined in (\ref{xie1}), we also introduce the
correlation 
function $\xiegal(r)$ due to collapsed halos which were able to cool and
to form 
galaxies. We estimate this contribution as follows. We assume that
within these 
halos the gas is totally ionized (by the radiation of the central stars
or QSO 
and by collisional ionization, due to shock-heating up to the virial 
temperature). Then, we can write (\ref{xie0}) as:
\beq
\xiegal(r) = \Fgalvol^2 \; (1+\Delta_c)^2 \; \xigal(r) = \Fgal^2 \;
\xigal(r)
\label{xiegal}
\eeq
where $\Fgalvol$ is the volume fraction (i.e. the filling factor)
occupied by 
these objects and $\Fgal = \Fgalvol (1+\Delta_c)$ is the fraction of
matter they 
contain. We note $\Delta_c(z)$ the mean density contrast of
just-virialized 
halos, given by the usual spherical model, while $\xigal(r)$ is the
matter 
correlation function associated with these halos. If we do not take into
account 
the density profile of these objects (i.e. we take a constant density
equal to 
$(1+\Delta_c) \rhoa$) we have for the clumping of the gas:
\beq
\Cgal = 1 + \xigal(r=0) = \frac{1+\Delta_c}{\Fgal} \gg 1 .
\label{Cgal}
\eeq
Then, in a fashion similar to (\ref{cutRd}) and (\ref{xiGM}) we write
the matter 
correlation function as:
\beq
r > \Rcool(z) : \hspace{0.2cm} \xigal(r) = (\Cgal-1) \; 
\frac{\xi(r)}{\xi(\Rcool)}
\label{xigal}
\eeq
and $\xigal(r) = (\Cgal-1)$ for $r<\Rcool$. Here we defined $\Rcool(z)$
as the 
virial radius of the smallest halos which are able to cool at redshift
$z$, as 
in Valageas \& Silk (1999a). Of course, our prescription for
$\xiegal(r)$ is 
only meant to provide a crude estimate of the contribution due to
galactic 
halos. Note that if we consider a specific density profile for the dark
matter 
halos we would only need to multiply $\Cgal$ by a numerical factor of
order 
unity, provided that the slope of the density in the inner regions is
shallower 
than $-3/2$ (otherwise $\Cgal$ is not finite). In addition, one should
take into 
account the additional collapse of the gas (while it cools) and the
effects of 
matter ejection through supernovae or stellar winds. In this article, we
shall 
restrict ourselves to the simple prescription (\ref{xigal}) which should
exhibit 
the right trend with $r$, keeping in mind that at small scales $r <
\Rcool$ we 
might underestimate the signal by a factor $\sim 3$. The comparison of
$\xiegal$ 
with $\xi_e$ allows us to see which of these two contributions (from
galactic 
halos or from the IGM) is dominant at a given scale. It also shows the 
characteristic scales associated with both processes.

\subsection{Power spectrum of secondary anisotropies}
\label{Power spectrum of secondary anisotropies}

Using the results of the previous sections, we can write the two-point 
correlation function $C(\theta)$ obtained in (\ref{Corr3}) as:
\beq
C(\theta) = \frac{2 \tau_0^2 v_0^2}{3 \; c^2} \int \d z \; w(z) 
\int_{R_{\theta}}^{\infty} \frac{\d r}{R_z} \; \frac{r \; \Pi_v[r (1+z)]
\; 
\xi_e(r)}{\sqrt{r^2-R_{\theta}^2}}
\label{Corr4}
\eeq
with:
\beq
w(z) = \frac{\d\chi}{\d z} \; \De(z) \; \left( \frac{H(z) f[\Om(z)]}{H_0 
f(\Omo)} \right)^2
\eeq
and:
\beq
R_z = \frac{c}{H_0} \; \frac{\De(z)}{1+z} .
\label{R0}
\eeq
In (\ref{Corr4}) we made the change of variable $r=| {\bf R}_{\theta} +
{\bf l} 
|$. The angular power-spectrum of these secondary anisotropies is given
by:
\beq
C_l = 2\pi \int \d (\cos \theta) \; P_l(\cos \theta) \; C(\theta)
\eeq
which for $l \gg 1$ can be approximated by:
\beq
l \gg 1 : \hspace{0.2cm} C_l = 2\pi \int_0^{\infty} \d\theta \; \theta
\; J_0(l 
\theta) \; C(\theta)
\label{Cl0}
\eeq
where $P_l$ are the Legendre polynomials and $J_0$ is the Bessel
function of 
order 0. Then, from (\ref{Corr4}) we obtain after integration over
$\theta$ (see 
Gradshteyn \& Ryzhik 1965, \S6.554.2, p.706):
\begin{eqnarray}
C_l & = & \frac{4 \pi \tau_0^2 v_0^2}{3 \; c^2 \; l} \int \d z \; w(z) 
\int_0^{\infty} \frac{\d r}{R_z} \; \frac{r}{R_z} \sin \left( \frac{l \;
r}{R_z} 
\right) \nonumber \\ \nonumber \\ & & \hspace{3.5cm} \times \;
\Pi_v[r(1+z)] \; 
\xi_e(r) .
\label{Cl1}
\end{eqnarray}
The expression (\ref{Cl1}) clearly shows that $C_l$ will exhibit a
strong 
decline at large $l$ for wavelengths which are smaller than the typical
scales 
of the free electron inhomogeneities ($R_z/l \ll R_{n_e}$) because of
the 
oscillatory term $\sin(l . r/R_z)$. On the other hand, for small $l$
(i.e. at 
large scales $R_z/l \gg R_{n_e}$ where the correlation of the electron 
distribution vanishes) we recover a white noise power-spectrum ($C_l$ is 
constant).

Finally, following Bruscoli et al. (2000) we define the quantity $S_l$
by:
\beq
S_l = \frac{1}{2\pi} \int_{l/2}^{3l/2} \d l \; l \; C_l \sim \lag
\frac{l (l+1) 
C_l}{2\pi} \rag_l.
\label{Sl0}
\eeq
This ``local average'' of $l (l+1) C_l/(2\pi)$ removes some of the
oscillations 
of $C_l$ at large $l$ and it allows us to see more clearly the drop at
large $l$ 
of the spectrum. It is also more convenient for numerical calculations
since it 
requires a lower resolution. From the relation (\ref{Cl1}) we obtain:
\begin{eqnarray}
S_l & = & \frac{2 \tau_0^2 v_0^2}{3 \; c^2} \int \d z \; w(z)
\int_0^{\infty} 
\frac{\d r}{R_z} \left( \cos \frac{l r}{2 R_z} - \cos \frac{3 l r}{2
R_z} 
\right) \nonumber \\ \nonumber \\ & & \hspace{3.5cm} \times \;
\Pi_v[r(1+z)] \; 
\xi_e(r) .
\label{Sl1}
\end{eqnarray}
Using (\ref{Cl0}) we can also write $S_l$ in terms of $C(\theta)$ as:
\beq
S_l = \int_0^{\infty} \frac{\d\theta}{\theta} \; C(\theta) \; \left[ 
\frac{3l\theta}{2} J_1\left( \frac{3l\theta}{2} \right) -
\frac{l\theta}{2} 
J_1\left( \frac{l\theta}{2} \right) \right] .
\label{Sl2}
\eeq

Note that the expressions (\ref{Corr4}), (\ref{Cl1}) and (\ref{Sl1}) are
quite 
general and do not depend on our model for the inhomogeneities of the
free 
electron distribution (which enters $\xi_e(r)$). They merely use the
linear 
evolution (\ref{v1}) of the velocity fluctuations and the approximation
that 
velocity and density fluctuations are uncorrelated. Moreover, it is more 
accurate (hence more convenient in terms of numerical resolution) to
compute 
$C_l$ and $S_l$ from (\ref{Cl1}) and (\ref{Sl1}) rather than from the
Fourier 
transform of the final angular correlation $C(\theta)$.

\section{Numerical results}
\label{Numerical results}

For the numerical calculations we consider an open CDM universe (OCDM)
with 
$\Om=0.3$, $\Ol=0$, $\Ob=0.03$, $H_0=60$ km/s/Mpc and $\sigma_8=0.77$.
These 
values are those we used in previous articles where we considered the
luminosity 
functions of galaxies (Valageas \& Schaeffer 1999), Lyman-$\alpha$
absorbers 
(Valageas et al. 1999), clusters (Valageas \& Schaeffer 2000) and
reionization 
by stars and quasars (Valageas \& Silk 1999a,b).

\subsection{Velocity fluctuations}
\label{Velocity fluctuations}

\begin{figure}[htb]

\centerline{\epsfxsize=8.2 cm \epsfysize=5.5 cm \epsfbox{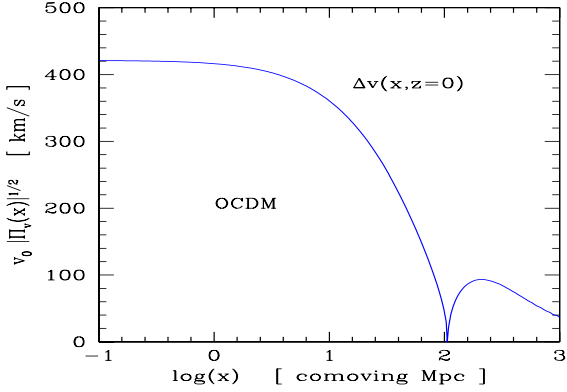}}

\caption{The magnitude of the velocity correlation function $\xi_v(x)$ at redshift $z=0$. The solid curve shows the quantity $v_0 \; |\Pi_v(x)|^{1/2}$, see (\ref{v1}). The feature at $x \simeq 100$ Mpc corresponds to the fact that $\Pi_v(x)$ becomes negative for $x \ge 100$ Mpc.}
\label{figvelo}

\end{figure}

First, we display in Fig.\ref{figvelo} the magnitude of the velocity correlation function $\xi_v(x)$ at $z=0$. It shows a plateau at small scales $x \ll R_{-1}$ (most of the power comes from $R_{-1}$ where the quantity $k P(k)$ is maximum) and it declines at large scales with oscillations. Thus, in our model velocity fluctations are of order $\Delta v \sim 400$ km/s at $z=0$ (and they roughly decrease as $(1+z)^{-1/2}$). They only arise from the linear growth of initial perturbations through gravitational interaction. In the actual universe, an additional source of velocity fluctuations (mainly at small scales) could be provided by other processes like the ejection of matter by supernovae or turbulence. If these velocity fluctuations are larger by a factor $\alpha$ than the value we use for $\Delta v$ shown in Fig.\ref{figvelo}, then on the corresponding scales we should roughly increase our predictions for the secondary CMB anistropies $C_l$ by a factor $(1+\alpha^2)$. We shall not investigate here this possibility but we can note that a significant effect would require rather large velocities.

\subsection{Free electron correlation function}
\label{Free electron correlation function}

The angular correlation $C(\theta)$ and the power-spectrum $C_l$
correspond to 
the sum over the line of sight of the fluctuations of the free electron
number 
density, up to the recombination redshift. Then, this integration over
redshift 
could blur some features of these density fluctuations. Hence, in order to
clarify 
the analysis it is interesting to consider first the real-space
correlation 
function $\xi_e(r;z)$ obtained for a given redshift $z$. This also
allows us to 
see the evolution with redshift of the free electron density
fluctuations.

In order to understand the physical origin of the signal we split up the 
correlation function into several parts. First, we consider the total 
contribution $\xi_e(r)$ from the IGM, as defined in (\ref{xie1}). Next,
we 
introduce $\xieu$ as the correlation function we get when we do not take
into 
account the correlations of ionized bubbles (subscript ``u'' for 
``uncorrelated''). That is, in (\ref{P2}) we set the term
$\xi_{x_1,x_2}(r)$ to 
0. Then, we define $\xieh$ as the signal produced by uncorrelated
bubbles into a 
``homogeneous'' IGM (subscript ``h'' for ``homogeneous''). That is, in 
(\ref{xie1}) we set the term $\xiIGM(r)$ to 0. Thus, $\xieh$ allows us
to see 
the contribution to $\xi_e$ due to the inhomogeneities of the free
electron 
number density produced by patchy reionization in distant bubbles. Then,
$\xieu$ 
shows by comparison with $\xieh$ the importance of the clumping of the
gas 
within the IGM. Finally, the difference between $\xieu$ and the total
signal 
$\xi_e(r)$ measures the effect of the correlations of these ionized
bubbles.

\begin{figure}[htb]

\centerline{\epsfxsize=8 cm \epsfysize=8 cm \epsfbox{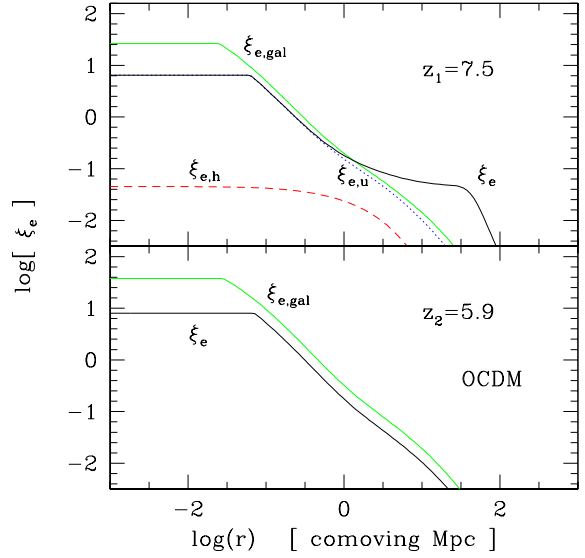}}

\caption{The real-space two-point correlation function $\xi_e(r)$ at two 
different redshifts. The solid line labeled $\xi_e$ (resp. $\xiegal$)
shows the 
contribution from the IGM (resp. from galactic halos). The curves
$\xieu$ and 
$\xieh$ correspond to the ``uncorrelated bubbles'' and ``homogeneous
IGM'' 
scenarios (see main text).}
\label{figXie2}

\end{figure}

We show our results in Fig.\ref{figXie2} for two different redshifts.
The upper 
panel at $z_1=7.5$ corresponds to a redshift slightly before
reionization (at 
$\zri=6.8$). First, we clearly see on $\xieh$ the contribution due to
patchy 
reionization within finite size bubbles. Thus, $\xieh$ is constant at
small 
scales below the characteristic size of the ionized bubbles ($\sim 0.5$
comoving 
Mpc) and it drops at large scales. Of course, since $\xieh$ only
corresponds to 
fluctuations in the ionized fraction (it does not take into account
matter 
density fluctuations) it is of the form $\xieh(r) = \lag x_{e,1} x_{e,2}
\rag(r) 
- \lag x_e \rag^2$. Since by definition we have $x_e \leq 1$ we get
$\xieh \leq 
1$. In fact, $\xieh$ never reaches unity because at large redshift the
filling 
factor $Q_{HII}$ of the ionized bubbles is much smaller than unity while
at low 
$z$ after reionization all the medium is reionized hence $\xieh(r) = 0$.
In the 
upper panel we have $Q_{HII} \simeq 0.9 < 1$ while in the lower panel
after 
reionization we have $\xieh(r) = 0$ hence the curve does not appear in
the 
figure. 

Next, $\xieu$ shows the influence of the clumping of the gas within the
IGM. 
Since $\CIGM > 1$ we no longer have the upper bound $\xieu \leq 1$ and
we can 
check in the figure that indeed we can have $\xieu > 1$. However, the
typical 
overdensities within the IGM are smaller than the density contrast
$\Delta_c(z)$ 
of just-virialized halos hence $\CIGM \leq 1+\Delta_c$, see the
expression 
(\ref{CIGM}). Thus, we obtain $\xieu \leq \Delta_c(z)$. We can check in
both 
panels in Fig.\ref{figXie2} that our results agree with this upper bound 
($\Delta_c(z) \sim 200$). In fact, $\xieu$ is significantly smaller
($\xieu \la 
10$) since most of the volume of the universe is filled by lower density
regions 
with $\rho \la \rhoa$. In agreement with (\ref{cutRd}) the correlation
function 
$\xieu(r)$ saturates at small scales below the damping scale $\Rd$ and
it 
follows the decrease at larger scales of the matter correlation
function. Since 
$\Rd$ is smaller than the typical sizes of the ionized bubbles the 
characteristic break of $\xieu(r)$ occurs at smaller scales than for
$\xieh$. In 
other words, the matter density fluctuations provide more small scale
power in 
relative terms than ionized bubbles. Moreover, since the clumping of the
gas is 
rather large, even at large redshifts (at $z\ga 10$ we already have
$\CIGM \ga 
10$, see Valageas \& Silk 1999a), we find that $\xieu \gg \xieh$ at all
scales.

However, the presence of the ionized bubbles is not totally blurred by
the 
superimposed matter density fluctuations. Indeed, we can see in the
upper panel 
that at large scales the actual correlation $\xi_e$ is much larger than
$\xieu$. 
This means that for $r \ga 1$ comoving Mpc the signal is dominated by
the 
cross-correlation of ionized bubbles. This arises from the correlations
of their 
central collapsed halos (which are the sites of formation of the central 
radiation source, either a galaxy or a QSO), see (\ref{P2}), which
correspond to 
very rare overdensities. Of course, by definition this effect appears at
scales 
larger than the typical size of the ionized bubbles. It provides excess
large 
scale power to $\xi_e$, as compared with $\xieu$, and it leads to a 
characteristic feature in the shape of the correlation function
$\xi_e(r)$. 
After reionization (lower panel) since there are no more ionized bubbles
this 
effect disappears and $\xi_e(r)$ is only governed by the fluctuations of
the 
matter density, hence $\xieu$ (not shown) becomes equal to $\xi_e$.

Finally, the curve $\xiegal$ shows the contribution from galactic halos.
We can 
see that the scale $\Rcool$ is somewhat smaller than the damping scale
$\Rd$ 
(this is related to the fact that the density of virialized halos is
larger by a 
factor $\Delta_c \sim 200$ than the typical IGM density) so that
$\xiegal$ 
saturates at smaller scales than $\xi_e$. Moreover, the clumping factor 
associated with these halos is larger than for the IGM (because of this 
difference between the typical densities) hence the plateau at small
scales of 
$\xiegal$ is higher than for $\xi_e$. From (\ref{xiegal}) and
(\ref{Cgal}) we 
can check that $\xiegal \leq \Fgal \Delta_c \leq \Delta_c$. We can see
in the 
figure that $\xiegal$ is somewhat smaller than this upper bound because
the 
fraction of matter enclosed within such halos is still low at these
redshifts: 
$\Fgal \sim 0.12$ at $z \sim 7$ (see also Fig.12 in Valageas \& Silk
1999a). We 
note that at larger scales ($r \sim 1$ Mpc) the contribution from the
IGM and 
``galaxies'' are of the same order while at very large scales ($r > 1$
Mpc) 
before reionization the signal is dominated by the IGM through the 
cross-correlation of ionized bubbles. We can expect to recover these
features in 
the integrated quantities $C(\theta)$ and $C_l$.

\subsection{Angular two-point correlation function}
\label{Angular two-point correlation function}

\begin{figure}[htb]

\centerline{\epsfxsize=8 cm \epsfysize=5.5 cm \epsfbox{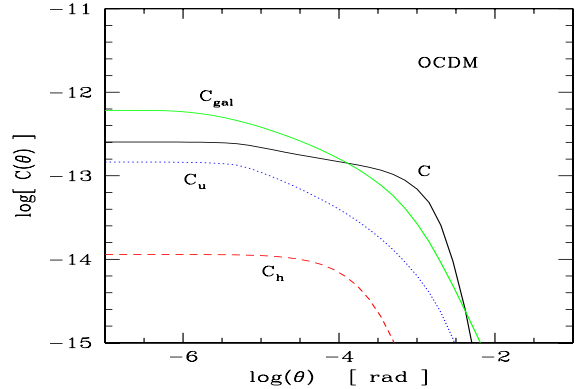}}

\caption{The angular two-point correlation function $C(\theta)$. The
solid line 
labeled $C$ shows the contribution from the IGM. The curve $\Cgal$
displays the 
contribution from galactic halos. The curves $\Cu$ and $\Ch$ correspond
to the 
``uncorrelated bubbles'' and ``homogeneous IGM'' scenarios (see main
text).}
\label{figCtheta}

\end{figure}

We show in Fig.\ref{figCtheta} our results for the angular two-point
correlation 
function $C(\theta)$. We can check that we recover the trends described
in 
Sect.\ref{Free electron correlation function}, which have not been
totally 
blurred by the integration over redshift.

First, we recover as in Fig.\ref{figXie2} the characteristic shape of
the 
contribution $\Ch$ from pure ionized bubbles: a plateau at small angular
scales 
$\theta \la 10^{-4}$ rad and a sharp drop at larger scales, beyond the
size of 
the bubbles. Note that a characteristic length scale $r$ is related to
the 
angular scale $\theta$ by $r \sim \theta c/H_0 \sim 5 \theta \times
10^3$ (in 
comoving Mpc). Hence $\theta \la 10^{-4}$ corresponds indeed to the
scale $r 
\sim 0.5$ Mpc seen in Fig.\ref{figXie2} (the knee of $\xieh(r)$). Then,
the 
correlation $\Cu$ shows a smoother shape, due to the power-law behaviour
of the 
real-space matter correlation function $\xi(r)$, with a break at a
smaller scale 
$\theta \sim 10^{-5.5}$ rad due to small scale matter density
fluctuations. 
Moreover, we have $\Cu > \Ch$, in agreement with Fig.\ref{figXie2}.
Next, the 
total signal $C(\theta)$ from the IGM is larger than $\Cu$, especially
at large 
angular scales $\theta \sim 10^{-3}$ rad, because of the
cross-correlation of 
ionized bubbles. Note that the integration along the line of sight
spreads the 
difference between $\Cu$ and $C$ over all angles $\theta$ (in particular
down to 
$\theta \rightarrow 0$) since a small angular separation $\theta$
corresponds to 
a large real-space distance $r$ at high $z$. This leads to a difference
with the 
real-space correlation $\xie(r)$ shown in Fig.\ref{figXie2} where we
found that 
at small scales (below the size of ionized bubbles) $\xieu$ is equal to
$\xie$. 
In a similar fashion, the integration over redshift also leads to
smoother 
curves $C(\theta)$. Finally, we can see that the contribution $\Cgal$
from 
galactic halos is larger than the signal from the IGM at small scales
$\theta 
\la 10^{-4}$ rad while it becomes smaller at larger scales, as expected
from 
Fig.\ref{figXie2}. However, on the whole the difference between both 
contributions is not very large.

\subsection{Power-spectra $C_l$ and $S_l$}
\label{Power-spectra Cl and Sl}

\begin{figure}[htb]

\centerline{\epsfxsize=8 cm \epsfysize=5.5 cm \epsfbox{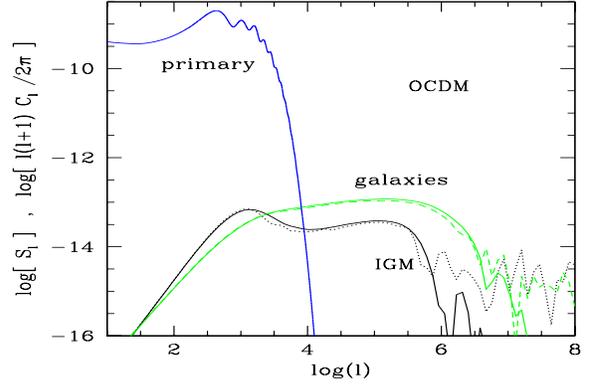}}

\caption{The power-spectra $C_l$ and $S_l$ of the secondary anisotropies
for the 
OCDM cosmology. The solid curves show the quantity $S_l$ for the
contributions 
from the IGM and from galactic halos. The dotted (resp. dashed) curve
with 
oscillations at large $l$ displays $l(l+1)C_l/(2\pi)$ for the IGM (resp. 
galactic halos). The upper curve labeled ``primary'' shows
$l(l+1)C_l/(2\pi)$ of the primary anisotropies for the same OCDM model (see text).}
\label{figClSl}

\end{figure}

We show in Fig.\ref{figClSl} the quantity $l(l+1)C_l/(2\pi)$ and the
``local 
average'' $S_l$, for the contributions from the IGM and from galactic
halos. 
We also plot for comparison the power spectrum of primary anisotropies 
calculated using CMBFAST (Seljak \& Zaldarriaga 1996).
At small $l$ both quantities $l(l+1)C_l/(2\pi)$ and $S_l$ are almost
identical 
since $C_l$ varies slowly with the wavenumber $l$. At large $l$ ($l \ga
10^5$) 
the power-spectrum $C_l$ exhibits an oscillatory behaviour (since it is
the 
Fourier transform of a function $C(\theta)$ which shows a sharp drop at
large 
angular scales) and a slow decline. In particular, at very large $l$ ($l
\ga 
10^7$) the oscillations of $C_l$ are not resolved by the numerical
calculation. 
On the other hand, $S_l$ becomes significantly different from
$l(l+1)C_l/(2\pi)$ 
as it shows a sharp decrease with $l$ and fewer oscillations. Of course,
this is 
due to the ``averaging procedure'' which enters the definition
(\ref{Sl0}) of 
$S_l$. At large $l$ the numerous oscillations of $C_l$ over the range 
$[l/2,3l/2]$ partially cancel out which leads to a stronger falloff for
$S_l$. 
Moreover, this ``averaging'' smoothes the behaviour of $S_l$ which shows
much 
weaker oscillations. This also allows us to resolve $S_l$ up to larger
$l$ (note 
that $S_l$ is not computed from $C_l$ but directly from the expression 
(\ref{Sl1})). This suggests that for observational purposes too, the
quantity 
$S_l$ may be more convenient as it should be more robust (i.e. require a
lower 
resolution) than $C_l$ at large $l$ and it shows more clearly the
transition to 
the large-$l$ regime by a sudden drop.

Note that for a correlation function $C(\theta) =
\exp[-(\theta/\theta_0)^2]$ 
with a Gaussian cutoff with a characteristic scale $\theta_0$ we get
from 
(\ref{Sl2}):
\beq
S_l = e^{-(l \theta_0)^2/16} - e^{-9(l \theta_0)^2/16} \sim e^{-(l 
\theta_0)^2/16}
\label{Slgauss}
\eeq
while (\ref{Cl0}) gives:
\beq
\frac{l^2 C_l}{2\pi} = \frac{1}{2} \; (l\theta_0)^2 \; e^{-(l
\theta_0)^2/4} .
\label{Clgauss}
\eeq
On the other hand, for an exponential cutoff $C(\theta) = 
\exp(-\theta/\theta_0)$ we obtain the same slow cutoff for both
quantities: $S_l 
\sim l^2 C_l \sim 1/(l \theta_0)$. Finally, for a correlation function
which is 
a top-hat (i.e. $C(\theta)=1$ for $\theta < \theta_0$ and $C(\theta)=0$
for 
$\theta > \theta_0$) we get:
\beq
S_l = J_0\left( \frac{l \theta_0}{2} \right) - J_0\left( \frac{3 l
\theta_0}{2} 
\right)
\eeq
and
\beq
\frac{l^2 C_l}{2\pi} = (l \theta_0) J_1(l \theta_0) .
\eeq
In this latter case, at large $l$ both $C_l$ and $S_l$ show an
oscillatory 
behaviour but we have $|S_l| \sim l^{-1/2}$ while $|l^2 C_l| \sim
l^{1/2}$, so 
that $S_l$ shows again a stronger decrease than $l(l+1)C_l/(2\pi)$.
Thus, these 
examples explicitly show that the shape of the large-$l$ tail of $S_l$
depends 
rather strongly on the details of the angular correlation function (and
even 
more so for $\Cl$). However, the ``local average'' $S_l$ usually shows a
drop 
beyond the characteristic wavenumber $1/\theta_0$ (while $|l^2 C_l|$ may
either 
decrease or grow in an oscillatory fashion).

The oscillatory behaviour of $C_l$ (hence the sharp drop of $S_l$)
appears at 
lower $l$ for the IGM than for galactic halos. This is due to the fact
that the 
correlations $\xie(r)$ and $C(\theta)$ show less small-scale power for
the IGM 
contribution than for the signal from galactic halos, as seen in 
Fig.\ref{figXie2} and Fig.\ref{figCtheta}. Indeed, as shown in
(\ref{Cl0}) and 
(\ref{Cl1}) a characteristic angle $\theta$ (resp. a physical length
$r$) 
translates into a characteristic wavenumber $l \sim 1/\theta$ (resp. $l
\sim 
R_z/r$). Hence, since the scales of the fluctuations of the free
electron number 
density are larger for the IGM than for the galactic halo component the 
power-spectra $C_l$ and $S_l$ of the IGM appear shifted towards smaller
$l$ with 
respect to the contribution from galaxies. At low $l$ we recover a white
noise 
spectrum ($C_l$ is constant, hence $S_l \propto l^2$) since this
corresponds to 
very large scales where the correlations of the electron distribution
are 
negligible. However, the slope of $\Sl$ we find at low $l$ for the
contribution 
from galactic halos is smaller because of the large-scale correlations
of these 
rare overdense objects.

We note that we clearly recover the main features of the correlation
$C(\theta)$ 
shown in Fig.\ref{figCtheta}. Thus, for the IGM we find that the
transition to 
the white noise part (the falloff at low $l$ with a slope $l^2$) occurs
at $l 
\sim 10^3$ which corresponds to the cutoff at large angles $\theta \sim
10^{-3}$ 
rad of the correlation $C(\theta)$ (see the strong knee in
Fig.\ref{figCtheta}). 
On the other hand, for the galactic halo contribution the transition to
the 
large-$l$ regime (marked by the drop of $S_l$) appears at $l \sim 10^6$
which 
corresponds to the scale $\theta \sim 10^{-6}$ rad below which
$\Cgal(\theta)$ 
saturates (hence to the smallest angular scale of the density
fluctuations). 
Note that additional power at smaller scales (due to substructures
within halos 
and to the subsequent collapse of baryons when they cool) would shift
this 
transition towards higher $l$.

\begin{figure}[htb]

\centerline{\epsfxsize=8 cm \epsfysize=5.5 cm \epsfbox{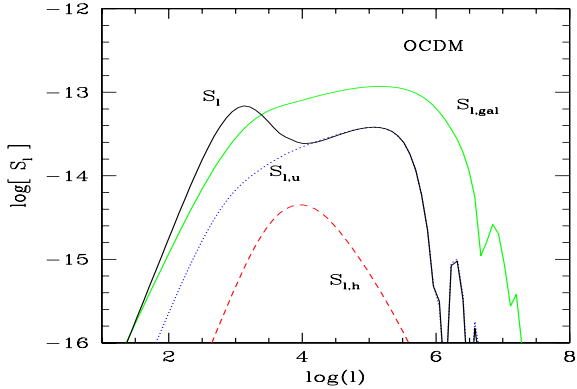}}

\caption{The power-spectrum $S_l$ of the secondary anisotropies for the
OCDM 
cosmology. The solid curve labeled $\Sl$ (resp. $\Slgal$) shows the
contribution 
from the IGM (resp. from galactic halos). The curves $\Slu$ and $\Slh$ 
correspond to the ``uncorrelated bubbles'' and ``homogeneous IGM''
scenarios.}
\label{figSl}

\end{figure}

In a fashion similar to Fig.\ref{figCtheta} we can split up the spectrum
$S_l$ 
into several components. This decomposition is shown in Fig.\ref{figSl}.
First, 
in agreement with Fig.\ref{figXie2} and Fig.\ref{figCtheta} we can see
that the 
contribution $\Slh$ from uncorrelated ionized bubbles within a uniform
medium is 
strongly peaked at $l \sim 10^4$ which corresponds to the typical size
of the 
ionized bubbles ($\theta_0 \sim 10^{-4}$ rad, $r \sim 0.5$ comoving
Mpc). At 
small wavenumber $(l \theta_0) \ll 1$ we recover a white noise spectrum
$S_l 
\sim (l\theta_0)^2$ while at larger $l$ we get a somewhat smoother
decrease than $l^{-2}$.

Next, in agreement with Fig.\ref{figCtheta} we can see that $\Slu$ is
larger 
than $\Slh$ and it shows a broader maximum, due to the additional power
at small 
and large scales provided by the density fluctuations within the IGM.
The 
comparison with the total signal $\Sl$ shows that this Fourier transform 
actually separates the various contributions to the power-spectrum.
Thus, for 
$l>10^4$ we have $\Sl=\Slu$ and the peak at $l \sim 10^{5.5}$
corresponds to the 
break at $\theta \sim 10^{-5.5}$ of the correlation $\Cu(\theta)$ (and
to the 
saturation at $r \sim 2 \times 10^{-2}$ Mpc of $\xieu(r)$). This
directly probes 
the small-scale fluctuations of the matter density field. Moreover, the
equality 
$\Sl=\Slu$ in this large-$l$ regime translates the fact that at small
scales $r 
< 1$ Mpc we had $\xie = \xieu$, as seen in Fig.\ref{figXie2}. Note that
in the 
angular space representation $C(\theta)$ the integration over redshift
along the 
line of sight destroys this feature as $C(\theta) > \Cu(\theta)$ for all
angles 
and one cannot recognize from the total signal $C$ the signature of this
small 
scale feature. Thus, the Fourier transform presents the strong advantage
to 
separate various physical processes as they act on different scales.
However, in 
this regime $l>10^4$ the signal should be dominated by the contribution
from 
galactic halos. 

Here we can note that in our calculations we modelled ionized regions as spherical bubbles while detailed numerical simulations show they can display a more complex morphology (e.g., Abel et al. 1999) as ionization fronts propagate more easily in voids. This means that the bell-shaped curve $\Slh$ in Fig.\ref{figSl} (which measures the contribution from patchy reionization, i.e. from the geometry of HII regions) underestimates the actual signal at large wavenumbers ($l > 10^4$) where we neglected the contribution from higher-order spherical harmonics (which provide some power on scales smaller than the typical radius $R_i$ of the ionized bubble). However, we can check in Fig.\ref{figSl} that even if we spread the maximum of the curve $\Slh$ up to wavenumbers $l$ ten times larger (i.e. the geometry of the ionization front displays significant power up to scales ten times smaller than $R_i$) our results remain unchanged. Indeed, most of the power in this range is due to the fluctuations of the matter density itself rather than to the geometry of HII regions. Moreover, the smallest scales displayed by the geometry of the ionization fronts are at least of the same order as the size of the smallest structures of the density field (from which they originate). Then, the factor $n_e^2$ in the correlation $C(\theta)$ ensures that if a large fraction of such clouds, enclosed within the distance $R_i$ to the radiation source, are ionized and if they show a density contrast $\delta \ga 10$, the signal is dominated on these scales by the fluctuations of the free electron density rather than by sheer geometrical patterns. Hence our results are not very sensitive to the approximation of spherical HII bubbles.

Then, the comparison of $\Sl$ with $\Slu$ shows that the peak at $l \sim
10^3$ 
comes from the correlations of ionized bubbles. This also corresponds to
the 
knee at $\theta \sim 10^{-3}$ rad of $C(\theta)$. In agreement with 
Fig.\ref{figCtheta} we find that at small wavenumbers $l \la 10^3$ the 
power-spectrum is dominated by the IGM contribution (i.e. the
correlations of 
ionized bubbles) while at larger $l$ most of the signal comes from
galactic 
halos which provide much small scale power. The peak at $l \sim 10^6$ of 
$\Slgal$ corresponds to the break at $\theta \sim 10^{-6}$ rad of
$\Cgal$. Thus, 
we see that the power-spectrum of the final signal mainly probes the
small-scale 
density fluctuations which give rise to galactic halos (and their
possible 
substructures) and the correlations at large scales of ionized bubbles.

Analytical studies of the power-spectrum $\Cl$ of these secondary CMB 
anisotropies have also been presented in some earlier works. Thus,
Gruzinov \& 
Hu (1998) consider the effect of patchy reionization within a uniform
medium, 
assuming a Gaussian falloff for the real-space correlation function
$\xie(r)$. 
This could arise from a Gaussian distribution of the size of the HII
regions. 
Then, as in (\ref{Slgauss}) and (\ref{Clgauss}) they obtain a Gaussian
cutoff 
for the angular correlation $C(\theta)$ and the power-spectra $\Sl$ and
$\Cl$. 
This corresponds to our curves $\Ch$ and $\Slh$ in Fig.\ref{figCtheta}
and in 
Fig.\ref{figSl}. Our predictions for this scenario are similar to their
results, 
but our spectrum $\Sl$ peaks at a larger $l$ ($l \sim 10^4$) than theirs
($l 
\sim 10^3$). This is due to the fact that in our model the comoving size
of HII 
regions is of order $0.5$ Mpc (see Fig.\ref{figXie2}) while they assume
a very 
large radius of $20$ Mpc for the bubbles. We can also note that our
cutoff at 
large $l$ is smoother than a Gaussian. Indeed, since observed luminosity 
functions usually show a simple exponential cutoff (and our results
match 
observations at low $z$) we can expect a shallow cutoff of the form 
$\exp(-r^{1/3})$ (because the volume of ionized bubbles is proportional
to the 
number of ionized atoms, hence to the luminosity of the source) and we
noticed 
above that a pure exponential cutoff already leads to a simple power-law
decline
of the spectrum $\Sl$ (as $1/l$).
Using a slightly more sophisticated model, Aghanim et al. (1996) calculate
the 
reionization from early formed quasars, deducing the statistic of the
ionized
bubbles size from the distribution of quasar luminosities. Their results 
predict that most of the power is at $l\sim 10^3$ and they are 
similar to a one-patch scenario with a bubble radius of $\sim 10$ Mpc,
except
for the high $l$ cutoff which is less steep, due to the smaller patches 
distribution.
Then, Knox et al. (1998) consider the effect 
of the correlations of these ionized bubbles. In agreement with our
results, 
they find that this leads to a much broader distribution of the
power-spectrum. 
Note that in our analysis we have split up the influence of matter
correlations 
into two processes: the cross-correlation of HII regions themselves
(through the 
correlation of the emitting sources), which provides additional power at
larger 
scales ($l<10^4$) than the size of these patches, and the fluctuations
of the 
matter density field within these bubbles, which builds power at smaller
scales 
($l>10^4$). Thus, the final signals $C(\theta)$ and $\Sl$ are the sum of
the 
contributions from various processes. The advantage of our approach is
that it 
provides a fully consistent description of these different scales, from
a model 
built to study the detailed reionization history of the universe which
has 
already been compared with observations for various aspects (e.g.,
galaxy 
luminosity function in Valageas \& Schaeffer 1999; X-ray emission from
clusters, 
galaxies and quasars in Valageas \& Schaeffer 2000).

\subsection{Redshift distribution}
\label{Redshift distribution}

The angular correlation $C(\theta)$ and the power-spectra $C_l$ and
$S_l$ 
correspond to an integration along the line of sight of the fluctuations
of the 
free electron number density. However, it would be interesting to see
the 
relative importance of the contributions from various redshifts to the
final 
signal. In particular, this would show whether these secondary CMB
anisotropies 
arise from a narrow range of redshifts close to reionization at
$\zri=6.8$ or 
from a more extended interval. Thus, we define the normalized quantity 
$\Ctheta(z)$ by:
\beq
\Ctheta(z) = \frac{2 \tau_0^2 v_0^2}{3 c^2 C(\theta)} \; w(z) 
\int_{R_{\theta}}^{\infty} \frac{\d r}{R_z} \; \frac{r \; \Pi_v[r (1+z)]
\; 
\xi_e(r)}{\sqrt{r^2-R_{\theta}^2}} .
\label{Cthetaz}
\eeq
It obeys the normalization condition:
\beq
\int_0^{\infty} \d z \; \Ctheta(z) = 1
\eeq
and $\Ctheta(z) \d z$ is the fraction of the final angular correlation
function 
which is generated in the redshift interval $\d z$. In a similar fashion
we 
define the quantities $\Cl(z)$ and $\Sl(z)$, from (\ref{Cl1}) and
(\ref{Sl1}), 
which are also normalized to unity. For these three redshift
distributions we 
alternatively consider the contributions from the IGM and from galactic
halos.

\subsubsection{IGM contribution}
\label{IGM contribution}

\begin{figure}[htb]

\centerline{\epsfxsize=8 cm \epsfysize=5.5 cm
\epsfbox{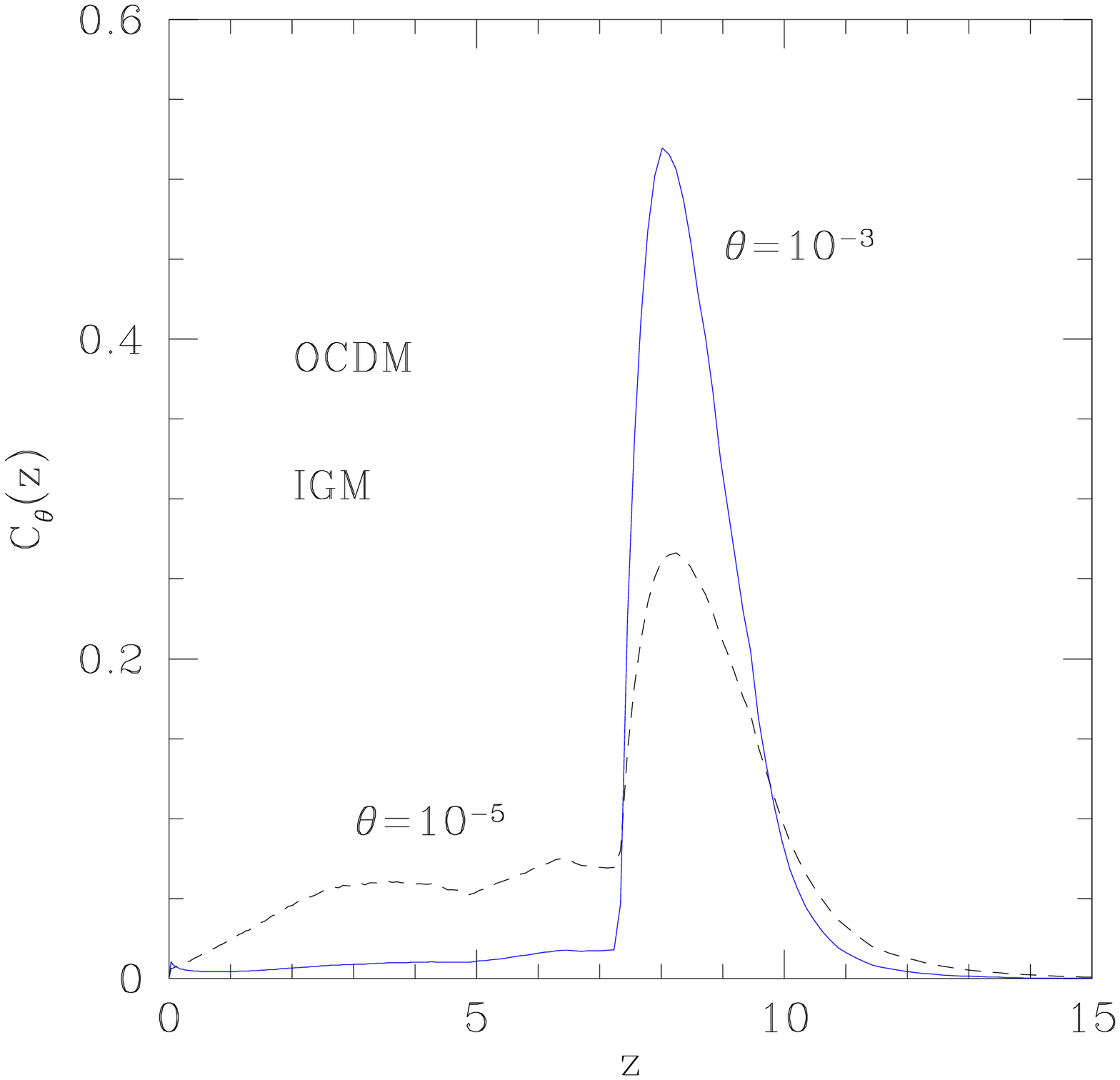}}

\caption{The redshift distribution $\Ctheta(z)$ of the angular
correlation 
(normalized to unity) from the IGM, for $\theta=10^{-3}$ rad (solid
line) and 
$\theta=10^{-5}$ rad (dashed line).}
\label{figCthetaz}

\end{figure}

\begin{figure}[htb]

\centerline{\epsfxsize=8 cm \epsfysize=5.5 cm \epsfbox{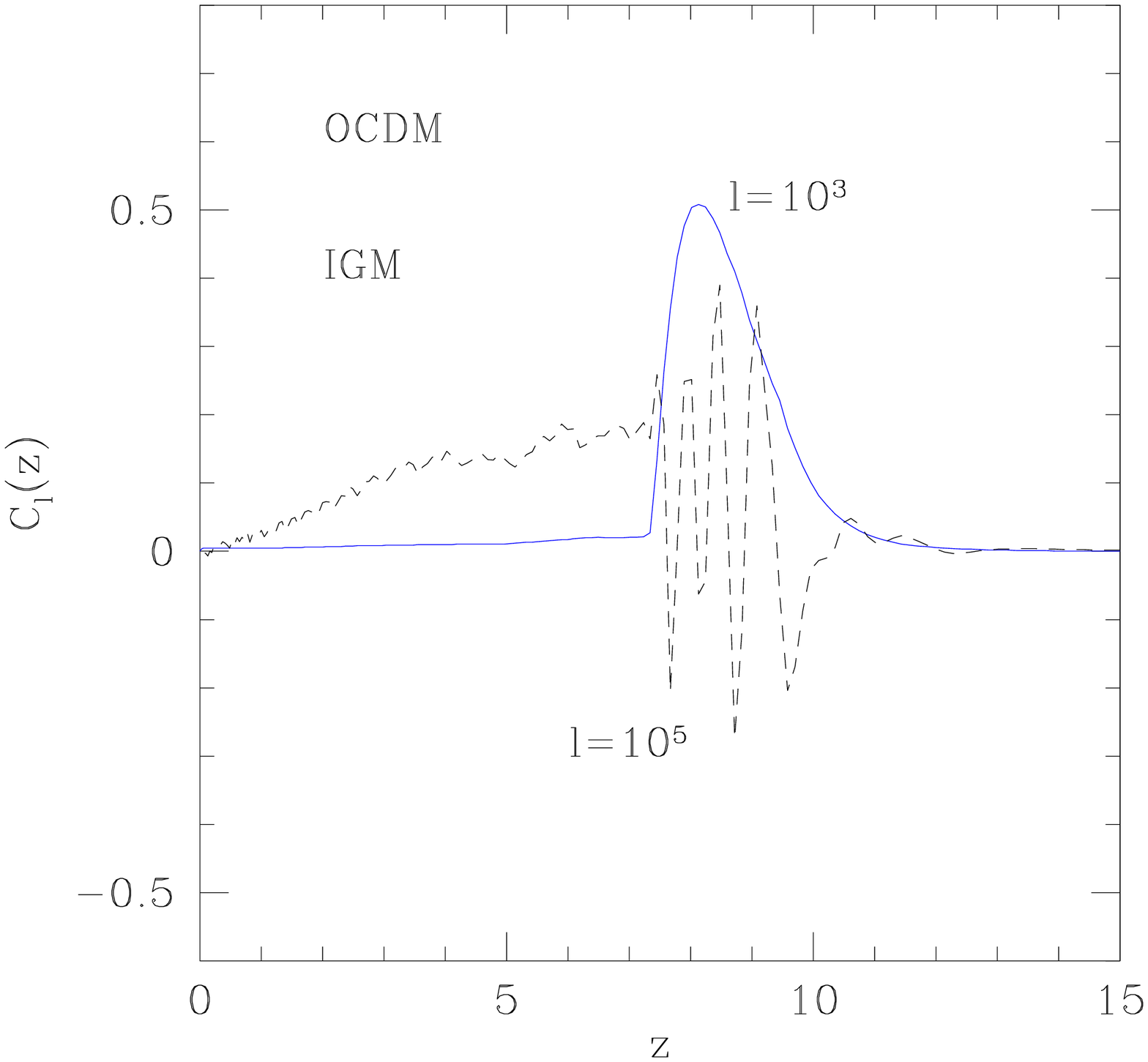}}

\caption{The redshift distribution $\Cl(z)$ of the power-spectrum $\Cl$ 
(normalized to unity) from the IGM, for $l=10^3$ (solid line) and
$l=10^5$ 
(dashed line).}
\label{figClz}

\end{figure}

First, we consider the contributions from the IGM to the secondary CMB 
anisotropies. We show our results for the redshift distribution
$\Ctheta(z)$ of 
the angular correlation in Fig.\ref{figCthetaz}, for two different
angular 
scales. We note that the contributions to the final signal $C(\theta)$
come from 
a rather large range of redshifts, typically $7.5 < z < 10$ (so that
$\delta z 
/\zri \sim 0.36$). There is a sharp cutoff at $z \simeq \zri$ since at
lower 
redshifts there are no more ionized bubbles. This drop is sharper for
larger 
angular scales, in agreement with Fig.\ref{figCtheta} and
Fig.\ref{figSl} where 
we noticed that large scales $\theta > 10^{-4}$ rad ($l<10^4$) are
dominated by 
the correlations of ionized bubbles. However, at small scales $\theta
\la 
10^{-5}$ there is a non-negligible tail at lower redshifts due to the
matter 
density fluctuations within the fully ionized IGM. Of course, smaller
scales 
also show a slightly more extended tail at high $z$ since at higher
redshift the 
typical size of ionized bubbles and the correlation length of the matter
density 
field were smaller, which damps the contribution to large angular scales 
$\theta$. Hence the redshift distribution of the angular correlation
$C(\theta)$ 
is somewhat broader for lower $\theta$ (which translates into the
smaller height 
of the maximum of $\Ctheta(z)$ in the figure since the curves are
normalized to 
unity).

\begin{figure}[htb]

\centerline{\epsfxsize=8 cm \epsfysize=5.5 cm \epsfbox{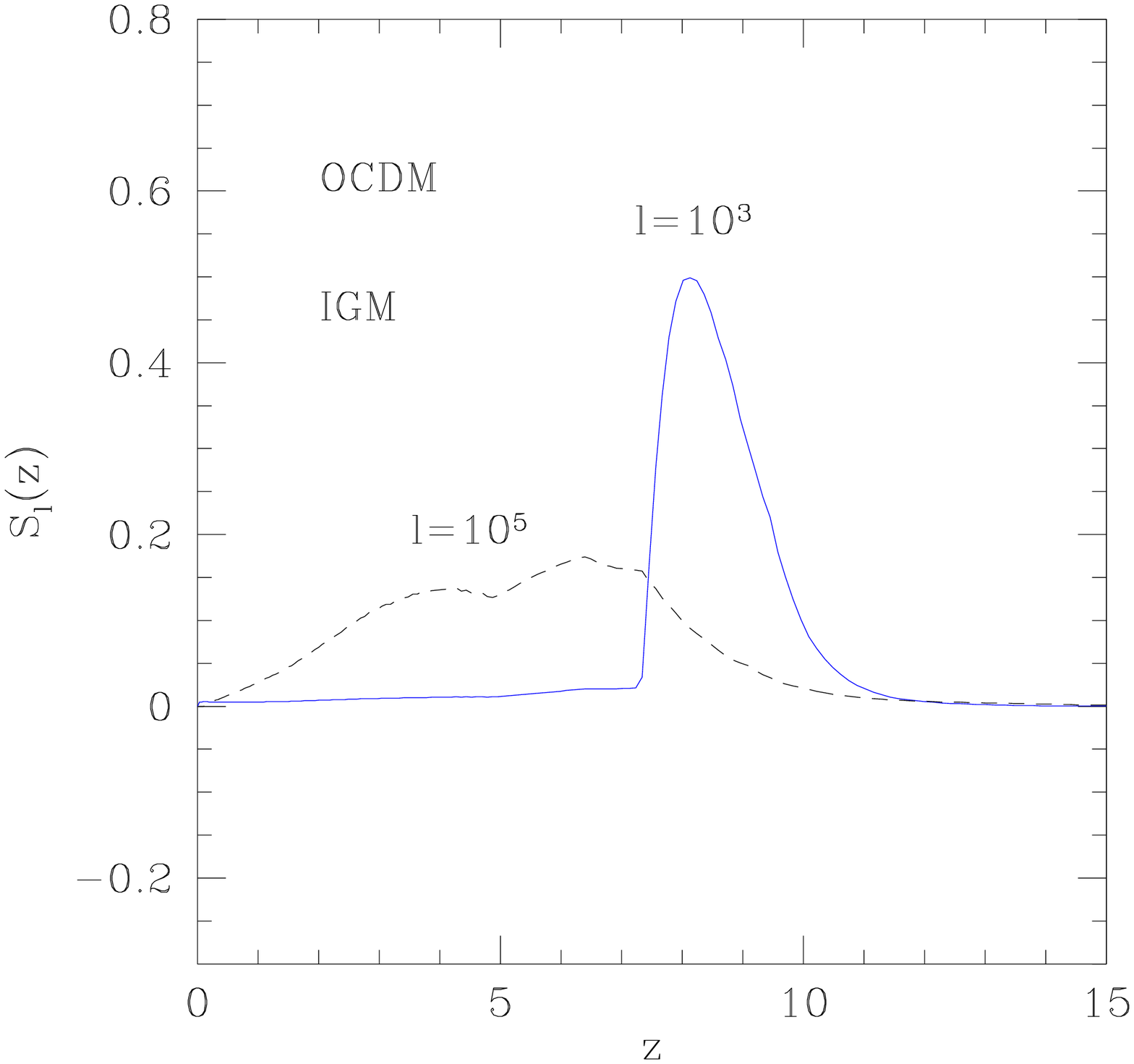}}

\caption{The redshift distribution $\Sl(z)$ of the power-spectrum $\Sl$ 
(normalized to unity) from the IGM, for $l=10^3$ (solid line) and
$l=10^5$ 
(dashed line).}
\label{figSlz}

\end{figure}

We display in Fig.\ref{figClz} the redshift distribution $\Cl(z)$ of the 
contribution from the IGM to the power-spectrum $\Cl$ for two different 
wavenumbers (normalized to unity). Of course, for $l=10^3$ we recover a
shape 
similar to the redshift distribution $\Ctheta(z)$ we obtained for 
$\theta=10^{-3}$ rad, since both quantities correspond to the same
scale. For 
$l=10^5$ the envelope of $\Cl$ agrees again with the shape we got in 
Fig.\ref{figCthetaz} for $\Ctheta(z)$ (with a tail at low $z$ due to
small-scale 
density fluctuations) but the distribution $\Cl(z)$ now shows several 
oscillations. This is due to the Fourier transform involved in the
definition of 
$\Cl$. Thus, the contributions from successive redshifts along the line
of sight 
almost cancel out. This agrees with the behaviour we obtained in 
Fig.\ref{figClSl}. Note that the oscillations occur before reionization:
they 
are due to the patchy pattern of reionization in HII bubbles with a size
larger 
than $R_z/l$. At lower redshift this feature disappears as there are no
more 
ionized regions to single out a large characteristic scale.

Finally, we display in Fig.\ref{figSlz} the redshift distribution
$\Sl(z)$ of 
the contribution from the IGM to the power-spectrum $\Sl$. As expected,
for 
$l=10^3$ we recover the results we obtained for the power-spectrum
$\Cl$. 
Indeed, as noticed in Fig.\ref{figClSl} for small wavenumbers there are
no 
oscillations since one probes scales which are larger or of the order of
the 
correlation lengths of the free electron distribution, so that $\Sl
\simeq 
l(l+1)\Cl/(2\pi)$. At larger $l$ some oscillations start to appear and
$\Sl(z)$ 
shows a different shape than $\Cl(z)$. In particular, the oscillations
of 
$\Sl(z)$ are much smoother and broader than for $\Cl(z)$ and they appear
at a 
larger wavenumber. Indeed, the ``local averaging'' over $l$ associated
with the 
procedure used to define $\Sl$ ``smoothes'' the contributions from
various 
scales. In particular, this allows us to see more clearly the redshift 
distribution associated with $l=10^5$ where there are no oscillations
yet. 
Moreover, it clearly shows that the large oscillations we obtained
shortly 
before reionization for $\Cl$ almost cancel out so that high redshifts
$z>\zri$ 
only provide a small contribution to the final signal. This leads to a
redshift 
distribution which is very different from the one obtained for smaller
$l$ which 
shows a sharp cutoff at $z\simeq \zri$. Thus, we find that for these
small 
scales the contributions to the power-spectrum $\Sl$ come from an
extended range 
of redshifts $2 < z <8$. As noticed in Fig.\ref{figSl}, we find again
that the 
use of the power-spectrum allows one to clearly see the various
processes 
associated with different scales, which are somewhat blurred in the
angular 
representation $C(\theta)$.

\subsubsection{Contribution from galactic halos}
\label{Contribution from galactic halos}

\begin{figure}[htb]

\centerline{\epsfxsize=8 cm \epsfysize=5.5 cm
\epsfbox{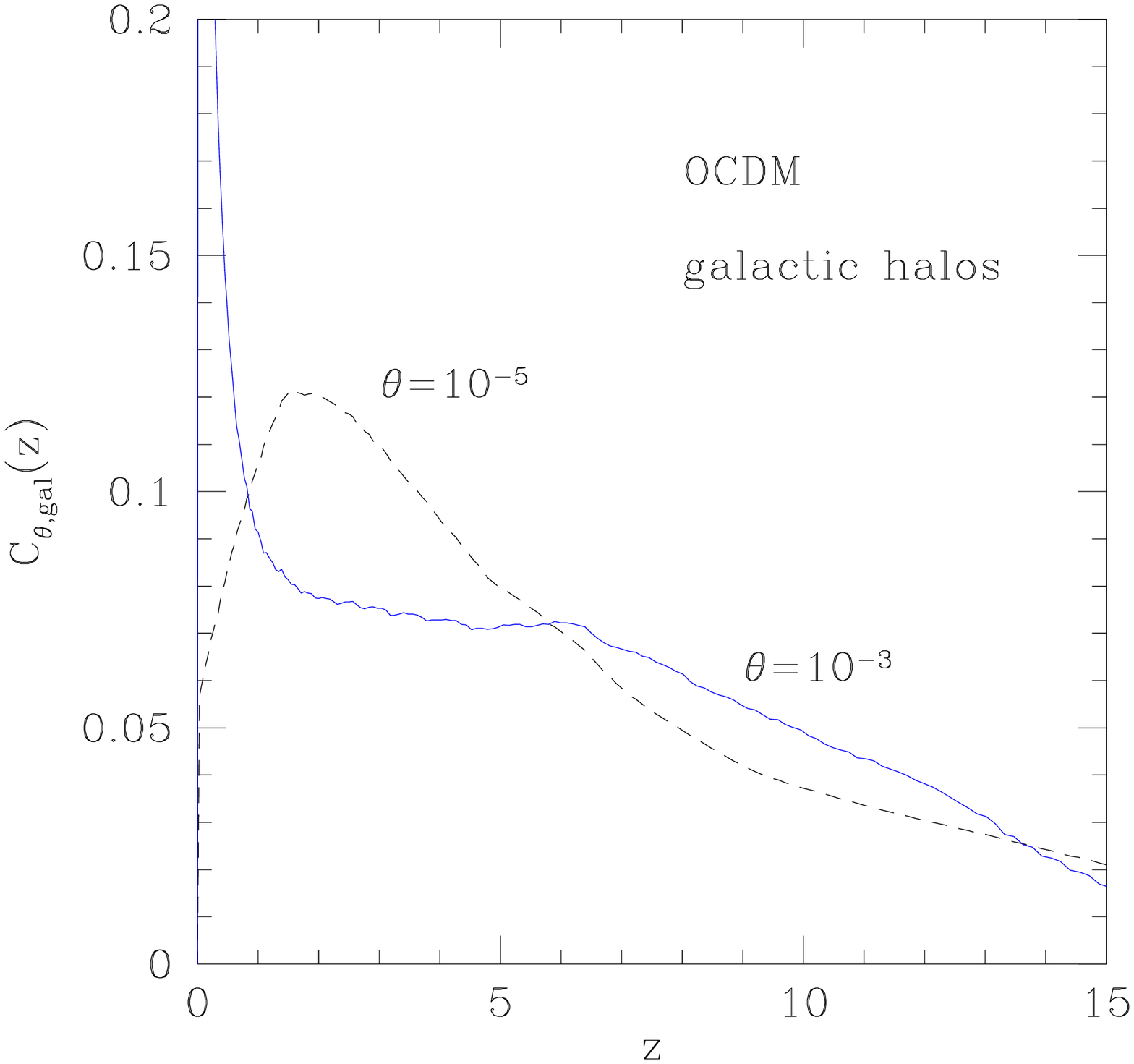}}

\caption{The redshift distribution $\Cthetagal(z)$ of the angular
correlation 
(normalized to unity) from galactic halos, for $\theta=10^{-3}$ rad
(solid line) 
and $\theta=10^{-5}$ rad (dashed line).}
\label{figCthetazcool}

\end{figure}

\begin{figure}[htb]

\centerline{\epsfxsize=8 cm \epsfysize=5.5 cm
\epsfbox{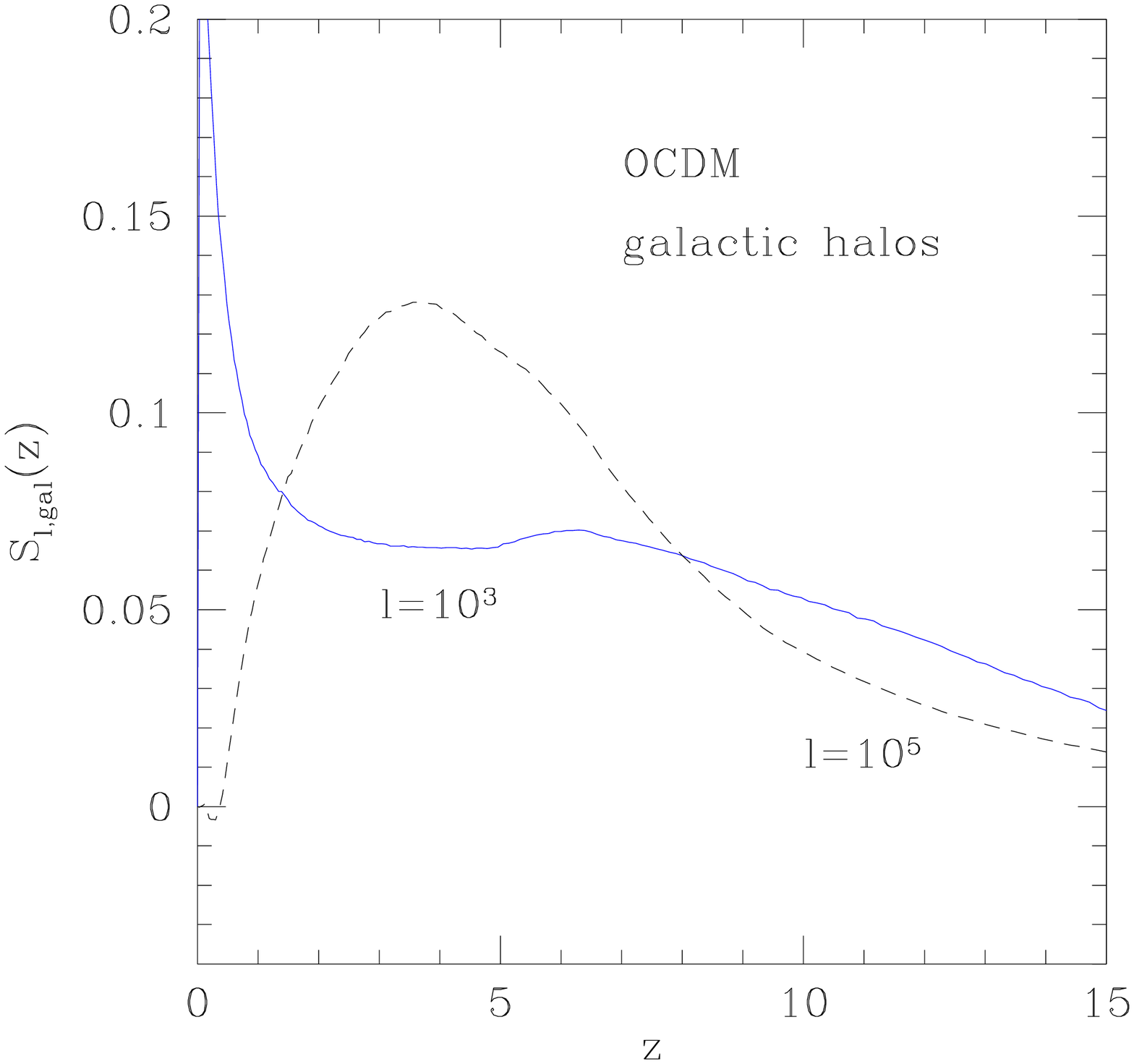}}

\caption{The redshift distribution $\Slgal(z)$ of the power-spectrum
$\Sl$ 
(normalized to unity) from galactic halos, for $l=10^3$ (solid line) and 
$l=10^5$ (dashed line).}
\label{figSlzcool}

\end{figure}

Now, we consider the redshift distribution of the contribution from
galactic 
halos to the CMB anisotropies. We show in Fig.\ref{figCthetazcool} our
results 
for the angular correlation function. First, we note that there is no
drop at 
the reionization redshift $\zri$. Indeed, the fact that ionized bubbles
suddenly 
overlap so that the signal from the patchy pattern of reionization
disappears 
does not affect the contribution from galactic halos. On the other hand, 
reionization does not lead to a sharp drop of the galaxy or quasar
multiplicity 
functions either since it does not imply a sudden increase of the IGM 
temperature and of the Jeans mass. Indeed, as seen in Valageas \& Silk
(1999a) 
most of the reheating of the universe occured earlier in a gradual
fashion so 
that the small increase of the IGM temperature at $\zri$ has no impact
on the 
population of radiation sources. Thus, the redshift distribution
$\Cthetagal(z)$ 
follows the growth of non-linear structures so that smaller redshifts
provide a 
larger contribution. This appears clearly for $\theta=10^{-3}$ rad where
most of 
the signal is generated at $z \la 1$ when the scale $r \sim 5$ Mpc
enters the 
non-linear regime. On the other hand, for the smaller angular scale 
$\theta=10^{-5}$ rad the contribution from very low $z$ becomes smaller
as the 
typical size of virialized objects becomes larger than the scale which 
corresponds to $\theta$. However, we may underestimate the power at low
$z$ 
because we neglected substructures within halos. Note that on these
small 
angular scales the secondary CMB anisotropies should be dominated by the 
contribution from galactic halos. Hence they arise from a very broad
range of 
redshifts (typically $0 < z <7$) which is not related to $\zri$.

We display in Fig.\ref{figSlzcool} the redshift distribution $\Slgal(z)$
of the 
power-spectrum $\Slgal$. We recover a behaviour similar to 
Fig.\ref{figCthetazcool}. In particular, note the large range of
redshifts which 
is probed by large wavenumbers $l \sim 10^5$. Higher $l$ which are
beyond the 
cutoff of the spectrum $\Sl$ show increasingly important oscillations.

\subsection{Quasars versus stars}
\label{Quasars versus stars}

In usual scenarios the universe is reionized by the radiation emitted by 
non-linear structures as collisional ionization is likely to be less
efficient 
(e.g., Madau 2000; Valageas \& Silk 1999b). There are two natural
sources of 
radiation in present cosmological models: stars and quasars. In
particular, in 
our model the universe is reionized when HII bubbles created by galaxies
and 
quasars overlap at $\zri$ (see Valageas \& Silk 1999a). The multiplicity 
functions we use for galaxies and QSOs are normalized to the
low-redshift 
universe ($z<4$) and are obtained in a consistent fashion (see also
Valageas \& 
Schaeffer 1999). Then, we find that the energy output provided by QSOs
is of the 
same order as the energy radiated by stars. However, the spatial
features of 
these two reionization processes may be different since one can expect
QSOs to 
create fewer but more extended HII bubbles, since quasars are not as
numerous as 
galaxies but their luminosity is much larger. Hence, the correlation
function 
$C(\theta)$ and the power-spectrum $C_l$ may show more large-scale power
for a 
quasar-driven reionization than for a galaxy-driven process. This is
quite 
interesting as it might allow one to discriminate both scenarios. Note
on the 
other hand that the reionization of helium is usually due to the
radiation 
emitted by quasars, as in our model. Indeed, stars have a black-body
spectrum 
which yields very few high energy photons while quasars exhibit a harder 
power-law spectrum over the relevant frequency range. However, as
pointed out by 
Tumlinson \& Shull (2000) population III metal-free stars have a harder
spectrum 
than typical low-$z$ stars, hence they might be able to ionize helium in 
addition to hydrogen. Thus, the relative importance of quasars and stars
is 
still an open problem. Unfortunately, we shall see below that the
observation of 
the secondary anisotropies of the CMB is unlikely to answer this
problem.

\begin{figure}[htb]

\centerline{\epsfxsize=8 cm \epsfysize=5.5 cm
\epsfbox{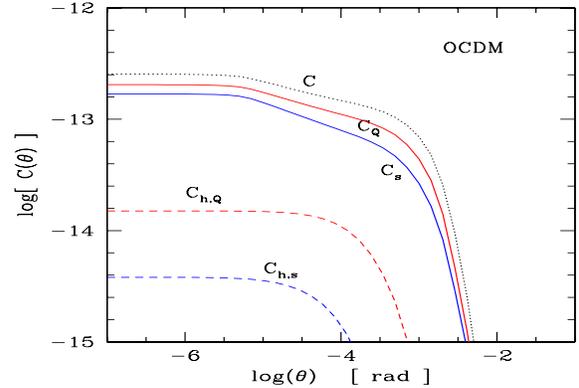}}

\caption{The angular two-point correlation functions $\Cs(\theta)$ and 
$\CQ(\theta)$ due to stars and quasars. The solid line labeled $\Cs$
(resp. 
$\CQ$) shows the contribution from the IGM when we only take into
account the 
ionized bubbles created by a central QSO (resp. by a central galaxy).
The curves 
$\Chs$ (resp. $\ChQ$) corresponds to the ``homogeneous IGM'' scenario of
ionized 
bubbles created by stellar radiation (resp. quasars) within a uniform
medium. 
The dotted line shows the total IGM correlation function $C(\theta)$ (it
is 
identical to the solid line $C$ in Fig.\ref{figCtheta}).}
\label{figCtheta0Q}

\end{figure}

Thus, we define $\Cs$ and $\Chs$ (resp. $\CQ$ and $\ChQ$) as the angular 
correlation functions we obtain for the total IGM signal and for the 
contribution due to patchy reionization through uncorrelated ionized
bubbles 
within a uniform IGM when we only count in our model the bubbles created
by 
stellar radiation (resp. quasar radiation). In other words, we use the 
reionization history described in the previous sections (see Valageas \&
Silk 
1999a) but to compute the CMB secondary anisotropies we only take into
account 
the bubbles associated with either one of the two available sources of
radiation 
(stars or quasars). This allows us to compare the importance of stars
and QSOs 
in our results (for the peculiar scenario of structure formation we
use). We 
show our results in Fig.\ref{figCtheta0Q}. First, we note that we
recover for 
both cases the main features described in Sect.\ref{Angular two-point 
correlation function} for the total signal. Then, as expected, the
comparison of 
$\Chs$ with $\ChQ$ shows that the characteristic scale of the ionized
bubbles 
associated with quasars is larger than for the HII regions produced by
galaxies. 
However, the difference is not very large (note that the radius only
scales as 
$L^{1/3}$, where $L$ is the source luminosity, and one has to integrate
over an 
extended population of sources and over redshift). Moreover, we find
that the 
total signals $\Cs$ and $\CQ$ are very close and they do not exhibit
different 
characteristic scales. Indeed, as we described in Sect.\ref{Angular
two-point 
correlation function} most of the power is provided by the small-scale
matter 
density fluctuations of the IGM and by the large-scale cross
correlations of 
ionized bubbles. Hence the typical size of the ionized bubbles cannot be
seen in 
the shape of the angular correlation $C(\theta)$. Besides, since quasars
and 
galaxies are drawn from the same population of collapsed halos they have
similar 
correlations hence the cross-correlations of their associated HII
regions are 
rather close (see Valageas et al. 2000 for a detailed study of the
correlation 
properties of these various objects).

\begin{figure}[htb]

\centerline{\epsfxsize=8 cm \epsfysize=5.5 cm \epsfbox{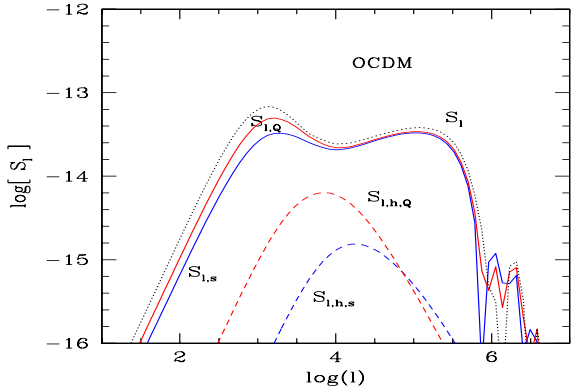}}

\caption{The power-spectra $\Sls$ and $\SlQ$ due to stars and quasars.
The 
dashed curve $\Slhs$ (resp. $\SlhQ$) correspond to the ``homogeneous
IGM'' 
scenario for ionized bubbles created by stellar radiation (resp.
quasars) within 
a uniform medium. The dotted line shows the total IGM spectrum $\Sl$ (it
is 
identical to the solid line $\Sl$ in Fig.\ref{figSl}).}
\label{figSl0Q}

\end{figure}

We show in Fig.\ref{figSl0Q} the power-spectra $\Sls$ and $\SlQ$ (as
well as 
$\Slhs$ and $\SlhQ$ for the ``homogeneous'' scenario) associated with
stars and 
quasars, which also correspond to the correlations displayed in 
Fig.\ref{figCtheta0Q}. We find again that the spectra $\Slhs$ and
$\SlhQ$ which 
directly probe the size of the HII regions exhibit two slightly
different scales 
for quasars and stars, in agreement with Fig.\ref{figCtheta0Q}. Thus,
$\Slhs$ 
peaks at $l \sim 10^{4.3}$ while $\SlhQ$ peaks at $l \sim 10^{3.8}$. The 
wavenumber associated with quasar-driven bubbles is smaller than for
stellar 
radiation since the size of the HII region is larger. However, we find
again 
that this signature is lost in the total power-spectra $\Sls$ and $\SlQ$
which 
are dominated at all scales by other processes (i.e. the correlations of
the 
matter density field itself). Thus, observations of the secondary
anisotropies 
of the CMB are unlikely to provide strong constraints on the size of the
ionized 
bubbles. Hence they cannot discriminate between both sources of
radiation (stars 
versus quasars).

Of course, this conclusion relies on the assumption that quasars are closely associated with galaxies. More precisely, our model is based on the usual scenario where QSOs correspond to massive black holes located in the nuclei of galaxies and powered by accretion (e.g., Rees 1984). Thus, an ``exotic'' model where quasars would not reside within massive virialized halos similar to galaxies might provide a different signature on the CMB. However, such a scenario is rather unlikely (e.g., in view of the energy requirements to power the quasars which favor large gravitational potential wells) and the standard model has been shown to agree reasonably well with numerous observations (e.g., the B-band luminosity functions and the X-ray emission, Valageas \& Schaeffer 2000). As we have shown above, on small scales ($l > 10^4$) the signal comes from the fluctuations of the density field within the IGM and from galactic halos, while the pattern of reionization plays a minor role. Hence our results in this range do not strongly depend on the clustering properties of quasars. On the other hand, on larger scales the kinetic SZ effect probes the spatial correlations of QSOs and in this sense it becomes more ``model-dependent''. However, we can be reasonably confident in our results as our model has already been checked against observations of the QSO multiplicity functions (e.g., Valageas \& Schaeffer 2000). Moreover, as shown in Fig.8 in Valageas et al. (2000) we also recover the observed behaviour with redshift of the correlation length associated with quasars. This means that any model which satisfies the same observational constraints (up to $z \la 5$) is likely to give analoguous results. Note that although the clustering properties of QSOs and galaxies as a whole are similar, since they are drawn from similar collapsed halos, the observed redshift-dependence of their correlation length is qualitatively different if one selects objects by a given luminosity threshold, due to the different behaviour of their mass-luminosity relations (see Valageas et al. 2000 for a detailed discussion).

\subsection{Dependence on cosmology}
\label{Dependence on cosmology}

\begin{figure}[htb]

\centerline{\epsfxsize=8 cm \epsfysize=5.5 cm \epsfbox{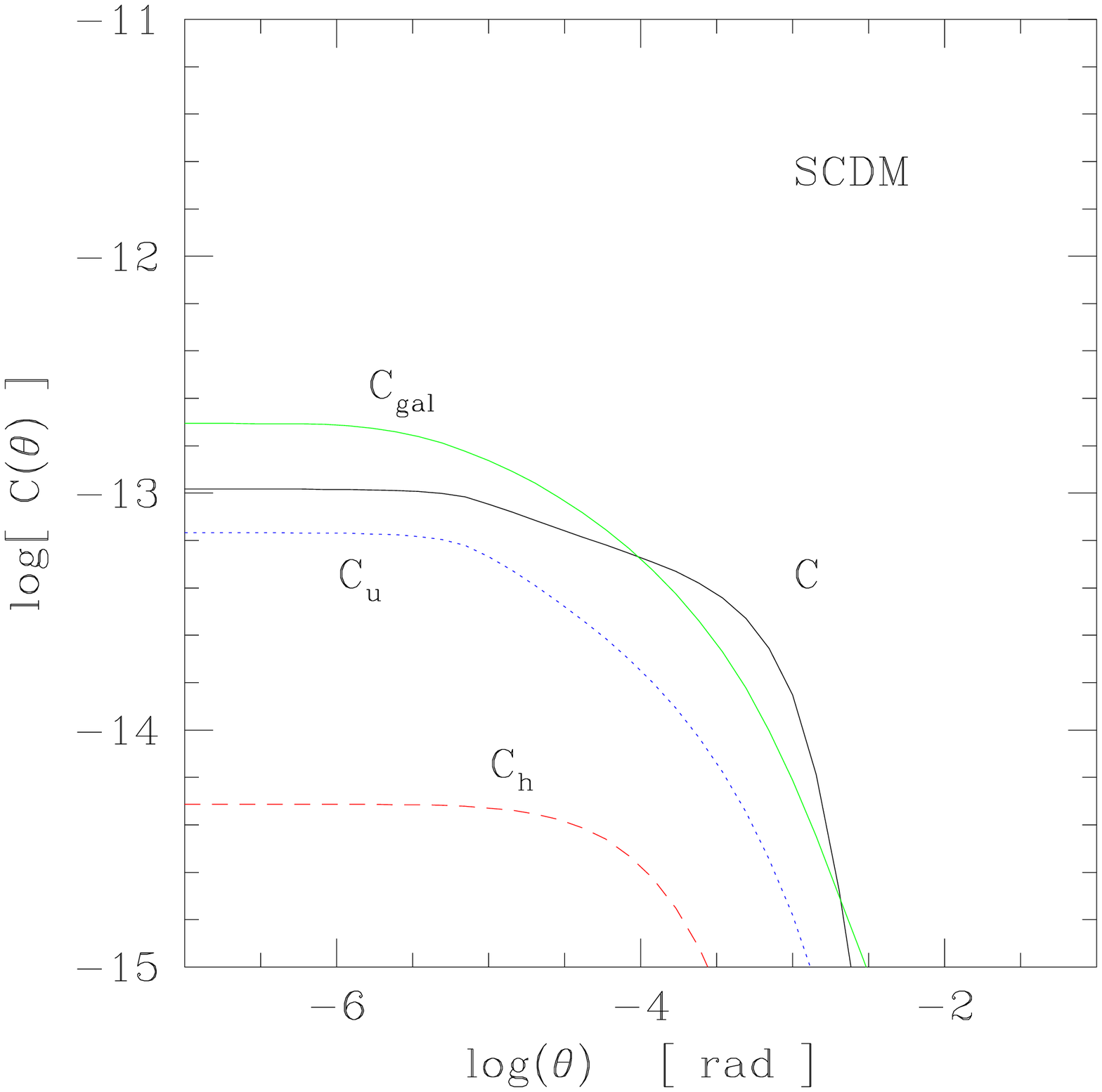}}

\caption{The angular two-point correlation function $C(\theta)$ for a
SCDM 
cosmology. The solid line labeled $C$ shows the contribution from the
IGM. The 
curve $\Cgal$ displays the contribution from galactic halos. The curves
$\Cu$ 
and $\Ch$ correspond to the ``uncorrelated bubbles'' and ``homogeneous
IGM'' 
scenarios, as in Fig.\ref{figCtheta}.}
\label{figCthetaO1}

\end{figure}

\begin{figure}[htb]

\centerline{\epsfxsize=8 cm \epsfysize=5.5 cm \epsfbox{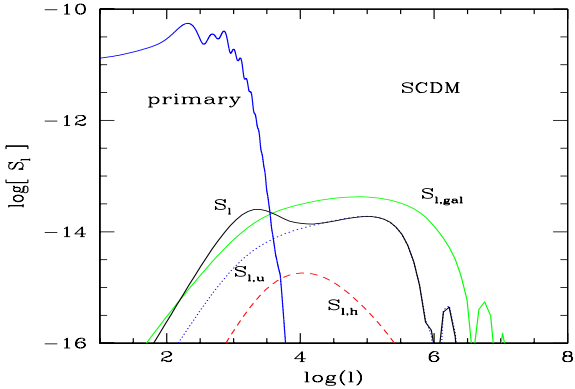}}

\caption{The power-spectrum $S_l$ of the secondary anisotropies for the
SCDM 
cosmology. The solid curve labeled $\Sl$ (resp. $\Slgal$) shows the
contribution 
from the IGM (resp. from galactic halos). The curves $\Slu$ and $\Slh$ 
correspond to the ``uncorrelated bubbles'' and ``homogeneous IGM''
scenarios, as 
in Fig.\ref{figSl}. The upper curve labeled ``primary'' shows
$l(l+1)C_l/(2\pi)$ 
of the primary anisotropies for the same SCDM model (see text).}
\label{figSlO1}

\end{figure}

Finally, in order to check whether our results strongly depend on the 
reionization history of the universe we also study the case of a
standard CDM 
cosmology (SCDM): $\Om=1$. We use the same model as in Valageas \& Silk
(1999a): 
$\Ob=0.04$, $H_0=60$ km/s/Mpc and $\sigma_8=0.5$. The physical processes 
included in the model are the same as for the open cosmology and the
galaxy and 
quasar multiplicity functions are again normalized to observations at
low $z$ 
(Valageas \& Schaeffer 1999; Valageas \& Silk 1999a). The reionization
redshift 
we get in this scenario is lower than previously: we now have
$\zri=5.6$.

We show in Fig.\ref{figCthetaO1} and in Fig.\ref{figSlO1} the angular 
correlation $C(\theta)$ and the power-spectrum $\Sl$ we obtain for this
critical 
density universe. First, we note that the amplitude of the secondary
distortions 
of the CMB is smaller than for the previous cosmology. This is due to
the 
smaller reionization redshift $\zri$. Indeed, the expression
(\ref{Corr4}) shows 
that:
\beq
C(\theta) \sim \delta z \; w(\zri) \sim \delta z \; (1+\zri)^{3/2}
\label{Czri}
\eeq
where $\delta z$ is the redshift interval during which reionization
occurs. The 
smaller reionization redshift is due to the smaller variance $\sigma_8$
and to 
the faster growth with redshift of density fluctuations (which implies
that for 
the same normalization at $z=0$ structure formation was less advanced at
high 
redshift for the SCDM cosmology). Next, we can check in the figures that
we 
recover the same features for the secondary anisotropies as for the open 
universe. In particular, we obtain the same shapes for all contributions
to the 
correlation functions and the relative importance of each process
remains 
unchanged. Hence all the points we discussed in the previous sections
for the 
low-density cosmology remain valid. This is not surprising since the
basic 
astrophysical model we use is the same. However, it shows that our
conclusions 
do not strongly depend on the details of the model (e.g., the
reionization 
redshift). 

For these reasons, we expect that for low-density flat 
models, such as the `cosmic concordance' $\Lambda$CDM model (Ostriker 
\& Steinhardt 1995; Krauss \& Turner 1995), the results should be
similar to the 
OCDM model, except for a shift of the features to lower $l$ (larger
angular 
scales) due to the different angular geometry.

\subsection{Comparison with numerical simulations}
\label{Comparison with numerical simulations}

Here, we briefly compare our results with other available studies. First, Benson et al. (2000) presented a semi-analytic model of galaxy formation to compute the kinetic SZ effect (within a $\Lambda$CDM cosmology). They roughly get the same amplitude $\Cl \sim 10^{-13}$ at $l \sim 10^4$ and they also find that at low wavenumbers ($l \la 10^3$) most of the power is provided by density fluctuations and the clustering of ionization sources. However, because of numerical resolution limitations they get a sharp drop at $l \sim 2 \times 10^4$ while we obtained a plateau for $\Cl$ up to $l \sim 10^6$. Note that our model is entirely analytic, although the scaling function $H(x)$ which enters the multiplicity functions in (\ref{Proba1}) is obtained from a fit to N-body simulations, so that we have no resolution limitations (we are simply limited by the approximations involved in our model).

The secondary anisotropies produced by inhomogeneous reionization which
we study in this article have also been computed by means of numerical
simulations in Bruscoli et al. (2000), Gnedin \& Jaffe (2000) and Springel et al. (2000), using different cosmologies and astrophysical models. These authors find a broad maximum for the power-spectra $\Cl$ and $\Sl$ of $\Cl \sim 10^{-13} - 10^{-12}$ around $l \sim 10^4$. In particular, Springel et al. (2000) get a slowly decreasing plateau down to $l \sim 500$ (below this scale they are limited by finite box size effects while both other numerical studies are restricted to $l \ga 10^4$). This behaviour agrees with our results (see the curve $\Slu$ in Fig.\ref{figSl}) since these authors use a simple toy model without including galaxy formation and radiative processes so that they miss the additional power due to the correlation of ionizing sources. Moreover, the drop we get for $\Cl$ at $l \la 300$, where we recover a white noise power
spectrum, is beyond the range of these numerical simulations. 

On small scales, Gnedin \& Jaffe (2000) (with the highest resolution) find a plateau at $\Cl \sim 10^{-13}$ which slowly decreases up to their resolution limit at $l \sim 10^6$. This again roughly agrees with our predictions, although we rather obtain a slight increase of the total power with $l$ in this range. Note that these numerical simulations are restricted to $z>4$ and these authors estimate the missing signal by a simple extrapolation (i.e. they multiply their output by a factor 1.25). However, as we discussed in Sect.\ref{Redshift distribution} the redshift distribution of the kinetic SZ effect depends on the angular scale one considers. Thus, large wavenumbers ($l \sim 10^5$), which probe high density fluctuations, are more sensitive to low $z$ than large scales ($l \sim 10^3$), which probe the inhomogeneous pattern of reionization and where most of the signal comes from epochs close to the reionization redshift $\zri$. This could explain the small difference between both predictions for the slope of this plateau. We can expect that with a higher resolution these authors would also recover a sharp cutoff at $l \ga 10^6$ for $\Sl$ (note that this drop is not readily apparent if one only computes the oscillatory spectrum $\Cl$, see Fig.\ref{figClSl}). These two behaviours are recovered by Bruscoli et al. (2000) at $l \sim 2 \times 10^5$, but this smaller value for the location of the transition might be due to the finite numerical resolution. Since this scale is directly related
to the size of virialized halos we can expect our result to be rather robust
(in fact we would even expect some power at slightly smaller scales due to the
collapse of baryons after they cool and to the substructures within halos, which would improve the agreement of our predictions with the results of Gnedin \& Jaffe 2000).

Bruscoli et al. (2000) also display the angular correlation function $C(\theta)$. It reaches a plateau $C(\theta) \sim 5 \times 10^{-12}$ for $\theta < 10^{-5}$ rad and it shows oscillations for $\theta > 2 \times 10^{-4}$ rad. Thus, the amplitude of the signal they get is larger than our predictions (as for $\Cl$). This could be due in part to their higher reionization redshift, see (\ref{Czri}). Moreover, Gnedin \& Jaffe (2000) argue that those authors overestimate the SZ effect by a factor 3-10 because of the uncorrected periodicity of the simulations. On the other hand, we obtain more large-scale power since in our model the cutoff of $C(\theta)$ only appears for $\theta \ga 10^{-3}$ rad. On these large scales, secondary CMB anisotropies are generated by the cross-correlations of ionized bubbles. Hence this difference between both predictions may also be related to our smaller value for $\zri$ since in our case at reionization structure formation is more advanced and the 
correlation length of the matter density field (hence of the radiation sources) 
is larger. Moreover, these large scales are not adequately resolved by these numerical simulations, as shown by their results for $\Cl$ which are restricted to $l \ga 10^4$. 
 
Finally, we note that the independent study by Gnedin \& Jaffe (2000) also finds that the signal is dominated by the contribution from high-density ionized regions rather than from the patchy pattern of reionization (for $l>10^4$). This agrees with our results. Note that we find in addition that the inhomogeneous pattern of reionization plays an important role at larger scales ($l \sim 10^3$) but this is beyond the range of these simulations. Thus, the agreement of our predictions with these various numerical studies, which use different cosmologies and astrophysical models, appears quite reasonable. Note that those numerical works do not include quasar formation models.

\section{Conclusion}
\label{Conclusion}

In this article, we have presented an analytic model (based on our
previous work 
which described structure formation processes and the reionization
history of 
the universe) which allows us to compute the secondary CMB anisotropies 
generated by the kinetic Sunyaev-Zel'dovich effect. This model includes
a 
consistent description of galaxies, quasars and matter density
fluctuations. 

We have found that {\it the contribution due to patchy reionization is
negligible 
except at very large scales} ($\theta \ga 10^{-3}$ rad) and small
wavenumbers ($l 
\la 10^3$). Over this range, which corresponds to scales larger than the
typical 
size of HII regions, the signal actually comes from the
cross-correlation of 
ionized bubbles, induced by the correlations of the rare radiation
sources. On 
smaller scales, the IGM contribution is governed by the fluctuations of
the 
matter density field itself. However, over this range the secondary
anisotropies 
should be dominated by the contribution from galactic halos, which are 
characterized by smaller scales than the IGM (and larger densities).
This leads 
to a cutoff of the power-spectrum $l(l+1)\Cl$ at a large wavenumber $l
\sim 
10^6$. On the other hand, at low wavenumbers $l < 10^3$ we recover a
white noise 
power-spectrum. This very extended range of wavenumbers $10^3 < l <
10^6$ is 
close to the limitations of current numerical simulations. Thus,
observations of 
{\it these secondary CMB anisotropies should mainly probe the correlation
properties 
of the underlying matter density field}, through the correlations of the
HII 
regions and the small-scale density fluctuations. We also found that the
``local 
average'' $\Sl$ of the power-spectrum should be a more convenient tool
than 
$\Cl$.

Some comments are in order about the detectability of the effects 
described in this paper. First, we notice that in the range 
$10^2 \la l \la 10^4$ the power
predicted by our model (relative to primary anisotropies) is comparable 
to the one found by Knox et al. (1998). Following their conclusions, we
infer that this signal, if not taken into account correctly, might 
introduce a small bias in the determination of cosmological parameters from 
future experiments like MAP (http://map.gsfc.nasa.gov) and particullarly Planck (http://astro.estec.esa.nl/SA-general/Projects/Planck/). Second, although the range of $l$ where our model produces most of the power 
($10^4\la l \la 10^6$) is likely to be out of reach for MAP and Planck, 
future mm-wavelength interferometers, such as ALMA (http://www.mma.nrao.edu) 
may have the right sensitivity ($\sim 2\mu$K rms for a $1'$ beam in one 
hour) and the right resolution ($< 2'$) to be able to measure such
a signal. Indeed, although the amplitude of the secondary anisotropies which we obtain is somewhat lower than the sensitivity of ALMA, a larger normalization of the power-spectrum (for the SCDM case we used $\sigma_8=0.5$ while a COBE normalization would give $\sigma_8 \sim 1.4$) or a larger reionization redshift would push the signal into the range of detectability.

We noticed that {\it the redshift distribution of the contributions to these 
secondary CMB anisotropies is rather broad}. Thus, for the angular
correlation 
from the IGM we get $7.5 < z < 10$ with a sharp cutoff at the
reionization 
redshift $\zri=6.8$, when the ``patchy pattern'' of hydrogen ionization 
disappears. However, some small-scale anisotropies are still produced at
lower 
redshifts.  The redshift distributions of the contributions from
galactic halos 
are even broader, we typically get $0<z<7$, and show no strong feature
at 
$\zri$. Since the total signal should be dominated by the contribution
from 
these collapsed objects for a large range of wavenumbers ($l \ga 10^4$)
this 
implies that one should not assume that most of the secondary CMB
anisotropies 
are generated during a small redshift interval $\delta z$ around the 
reionization redshift $\zri$.

Next, as expected we have found that within our scenario ionized bubbles 
produced by quasars are larger than those built by galaxies. This
implies that 
the ``patchy patterns'' of the HII regions associated with QSOs and
stars are 
different. However, since the total signal is dominated by the
correlations of 
the matter density field, and not by the size of the ionized bubbles, it
is 
similar for both radiation sources (which also have similar correlation 
properties). Hence, unfortunately {\it one cannot distinguish a quasar-driven 
reionization process from a galaxy-driven reionization history}, using
the CMB 
anisotropies.

Finally, we have checked that our predictions apply both for an open
cosmology 
and for a critical density universe. Thus, our conclusions do not depend
on the 
cosmological scenario and can be extended to low-density flat
models. 
However, the amplitude of the anisotropies is larger for 
the low-density universe because of the higher reionization redshift.

\end{document}